\documentclass[fleqn,usenatbib]{mnras}

\usepackage{newtxtext,newtxmath}

\usepackage{amsmath}
\usepackage{amssymb}
\usepackage{placeins}
\newcommand{\citeo}{\href{\#cite.10.1093/mnras/stae1399}{O24}}
\usepackage{lettrine}

\usepackage{xcolor}
\definecolor{darkgreen}{HTML}{4fb832}
\newcommand{\red}[1]{\textcolor{black}{#1}}
\newcommand{\green}[1]{\textcolor{black}{#1}}
\newcommand{\orange}[1]{\textcolor{black}{#1}}
\newcommand{\cyan}[1]{\textcolor{black}{#1}}
\newcommand{\newred}[1]{\textcolor{black}{#1}}

\usepackage{subcaption}
\usepackage{caption}

\usepackage[T1]{fontenc}

\DeclareRobustCommand{\VAN}[3]{#2}
\let\VANthebibliography\thebibliography
\def\thebibliography{\DeclareRobustCommand{\VAN}[3]{##3}\VANthebibliography}

\usepackage{graphicx}	
\usepackage{amsmath}	

\title[Cosmic dipole tensions]{Cosmic dipole tensions: confronting the cosmic microwave background with infrared and radio populations of cosmological sources}

\author[M. Land-Strykowski et al.]{
Mali Land-Strykowski,$^{1}$\thanks{E-mail: mali.land-strykowski@sydney.edu.au (MLS)}
Geraint F. Lewis$^{1}$
and Tara Murphy$^{1}$
\\
$^{1}$Sydney Institute for Astronomy, School of Physics, A28, The University of Sydney, NSW 2006, Australia
}

\date{Accepted XXX. Received YYY; in original form ZZZ}

\pubyear{\the\year{}}

\begin{document}
\label{firstpage}
\pagerange{\pageref{firstpage}--\pageref{lastpage}}
\maketitle

\begin{abstract}
The cosmic dipole measured in surveys of cosmologically distant sources is generally \green{found to be} in disagreement with the kinematic expectation of the Cosmic Microwave Background (CMB). This discrepancy represents severe tension with the Cosmological Principle and challenges the standard model of cosmology. Here, we present a Bayesian analysis of the tension between datasets used to measure the cosmic dipole. We examine the NRAO VLA Sky Survey (NVSS), the Rapid ASKAP Continuum Survey (RACS) and the Wide-field Infrared Survey Explorer catalogue (CatWISE), and jointly analyse them with the \textit{Planck} observations of the CMB. \red{Under the kinematic interpretation,} we find that \textit{Planck} is in severe tension with CatWISE \orange{above} $5\sigma$, \red{strong tension with RACS, and moderate tension with NVSS. Moreover, the strong concordance between CatWISE and NVSS suggests that their dipoles arise from a common astrophysical signal. Conversely, the high discordance between RACS and both CatWISE and NVSS indicates a possible systematic difference in the \orange{RACS} catalogue itself.} Whilst the tension between \textit{Planck} and infrared-selected quasars is already significant, the question of whether or not the dipole in \red{individual} radio surveys adds to the challenge against the standard model is yet to be seen. We estimate that $\mathcal{O}(10^6)$ radio sources are required to measure the tension to a significance of $5\sigma$. Therefore, in light of the upcoming SKA radio surveys, we are on the cusp of disentangling the anomaly of the cosmic dipole.
\end{abstract}




\section{Introduction}
\lettrine{T}{he} Cosmological Principle (CP) that underpins our modern cosmological view postulates that the Universe is statistically homogeneous and isotropic at sufficiently large spatial scales. The standard concordance model of cosmology, Lambda Cold Dark Matter ($\Lambda$CDM), inherently assumes the CP in the derivation of the Friedmann--Lemaître--Robertson--Walker (FLRW) spacetime \citep[see e.g.,][]{1968JMP.....9.1344E,ELLIS1983487,1995ApJ...443....1S,Clarkson_2010,CLARKSON2012682}. \red{Such an assumption is supported by the smoothness of the Cosmic Microwave Background (CMB), whereby thermal equilibrium resulted via causal contact in the early Universe \citep[for a recent review see][]{Vazquez_2020}.}

Imprinted in the CMB is a large-scale anisotropy, a temperature dipole of order $\Delta T/T\!\approx\!10^{-3}$. This is generally interpreted as being a kinematic effect from our heliocentric motion relative to the frame in which the CMB is maximally isotropic \citep{PhysRevLett.18.1065,PhysRev.174.2168,doi:10.1098/rsta.2011.0289}. Such a motion has been constrained to $v_\text{CMB}\!=\!369.82\pm0.11$~km~$\text{s}^{-1}$ towards $(l,b)\!=\!(264.021\pm0.011^\circ, 48.253\pm0.005^\circ)$ \citep{2020A&A...641A...1P}, in general alignment with the Great Attractor, an over-density of matter in the local Universe \citep[see, e.g.][]{1989Natur.342..251R,1990ApJ...354...18B,1991Natur.350..391D}.


\red{In $\Lambda$CDM, a primordial dipole in the CMB should be insignificant compared to the kinematic dipole when adopting typical departures from the Hubble flow \citep{Spergel_2003}. Since the density fluctuations traced by the CMB anisotropies grew via gravitational instability to form the large-scale structure in the Universe \citep[][]{2010fimv.book..267S}, any dipole in our observations of cosmologically distant matter should be kinematic and consistent with that seen in the CMB.}

If the kinematic interpretation is correct, this consistency check is a direct test of the CP. Such a test using cosmologically distant sources was first proposed by \citet{1984MNRAS.206..377E}. Given a flux density limited survey, our motion induces a dipole in the source counts by boosting sources above and below the survey limits due to relativistic Doppler shift and aberration. Specifically, given a survey of sources with intrinsic spectral index $\alpha$ with $S\!\propto\!f^{-\alpha}$ and limiting flux density threshold $S_*$, the integral source count per unit solid angle changes as $dN(>\!S_*)/d\Omega\!\propto\!S_*^{-x}$, where $x$ is the slope of the integral source counts. The resulting dipole anisotropy $\Delta N/N\!=\!\mathbf{d}\cdot\hat{\mathbf{n}}$ with velocity $\beta\!=\!v/c$ then has an amplitude of
\begin{align}
    \mathcal{D}=(2+x[1+\alpha])\beta.\label{eq:dipole-amplitude}
\end{align}

In the decades following, the \citet{1984MNRAS.206..377E} test has been applied to a \orange{large} number of surveys. The resulting \orange{source count} dipole has, in general, had a direction that aligns with the CMB dipole but an amplitude that significantly exceeds the kinematic expectation \citep[for a recent review see, e.g.][]{PEEBLES2022169159,2022JHEAp..34...49A,KumarAluri_2023,2025NatRP...7...68S}. This discrepancy is in severe tension with the CP to the extent that the standard cosmological model has been challenged. \red{In the literature, tension has been revealed by frequentist statistical comparisons of the dipole \orange{inferred in surveys} to the CMB expectation \citep[see, e.g.][]{Secrest_2022} and by \orange{performing} Bayesian model comparisons on survey data to the kinematic CMB hypothesis (see, e.g. \citealt{10.1093/mnras/stae1399}, hereafter \citeo). However, in a Bayesian framework, no study has quantified the tension between these surveys.}

In this paper, we present a Bayesian analysis that quantifies the tension between datasets used to measure the dipole under the kinematic interpretation. We investigate the tension between CMB observational data, namely the \textit{Planck} Public Data Release 3 \citep{2020ipac.data.I558P}, and three surveys used to measure the dipole in recent literature\orange{:} the National Radio Astronomy Observatory Very Large Array Sky Survey \citep[NVSS;][]{Condon_1998}, the Rapid Australian Square Kilometre Array Pathfinder Continuum Survey \citep[RACS;][]{McConnell_2020}, and quasars derived from the Wide-field Infrared Survey Explorer catalogue \citep[CatWISE;][]{Wright_2010}. In Section~\ref{sec:tension-explained}, we outline the theory of Bayesian tensions. We introduce our samples and data processing in Section~\ref{sec:nvss-and-racs} and detail our joint analysis approach in Section~\ref{sec:nested-sampling-approach}. In Section~\ref{sec:results}, we present our results and we discuss them in Section~\ref{sec:discussion}, where we conclude with our principal findings.

\section{Background}\label{sec:tension-explained}
\subsection{Bayes's Theorem}
We use a Bayesian framework for parameter estimation and evidence computation, in line with recent analyses of the cosmic dipole \citep[see, e.g.][\citeo]{10.1093/mnras/stad2322,Wagenveld_2023,10.1093/mnras/stad3706}. Here, we employ Bayes's Theorem with the notation 
\begin{align}
    P(\theta|D)=\frac{P(D|\theta)P(\theta)}{P(D)}~~~\Leftrightarrow~~~\mathcal{P}_D(\theta)=\frac{\mathcal{L}_D(\theta)\pi(\theta)}{\mathcal{Z}_D},\label{eq:bayes-theorem}
\end{align}
where dataset $D$ has posterior $\mathcal{P}_D$, likelihood $\mathcal{L}_D$, prior $\pi$ and evidence $\mathcal{Z}_D$, for some set of parameters $\theta$. 
$\mathcal{Z}_D$ is the marginal likelihood integrated over the set of parameters, whereby
\begin{align}
    \mathcal{Z}_D=\int\mathcal{L}_D(\theta)\pi(\theta)~d\theta.
\end{align}
If two datasets $A$ and $B$ are independent, the evidence can be combined at the likelihood level, such that $\mathcal{L}_{AB}=\mathcal{L}_A\mathcal{L}_B$, and
\begin{align}
    \mathcal{Z}_{AB}=\int\mathcal{L}_{AB}(\theta)\pi(\theta)~d\theta,~\text{where}~\mathcal{Z}_{AB}\neq\mathcal{Z}_A\mathcal{Z}_B.
\end{align}

\subsection{The Bayesian Evidence Ratio}\label{subsec:independant-datasets}
The Bayesian evidence ratio, or Bayes ratio $R$, quantifies the consistency between two or more datasets \citep{PhysRevD.73.067302}. It was constructed in the context of cosmological tensions and is defined for independent datasets, such that
\begin{align}
    R=\frac{\mathcal{Z}_{AB}}{\mathcal{Z}_A\mathcal{Z}_B},\label{eq:R}
\end{align}
where $\mathcal{Z}_A$ and $\mathcal{Z}_B$ are the probabilities of observing the individual datasets $A$ and $B$ under the common model, respectively, and $\mathcal{Z}_{AB}$ is the probability of observing the datasets combined. If the probability of observing the datasets combined is greater than the probability of observing the datasets individually ($\mathcal{Z}_{AB}\!\gg\!\mathcal{Z}_A\mathcal{Z}_B$) this implies that the datasets are consistent and $R\!\gg\!1$ (and vice versa for tension with $R\!\ll\!1$). Consistency occurs when combining the datasets substantially increases the evidence beyond that if they were analysed individually. The datasets in the numerator of Equation \ref{eq:R} must both be described by the same parameter values in the model, whereas those of the denominator can vary. The addition of unconstrained or nuisance parameters does not affect the inference of $R$, since only the shared, constrained parameters contribute to the Bayesian evidence.

The $R$ statistic can also be interpreted directly through Bayes's Theorem \citep{10.1093/mnras/stt008,PhysRevD.99.043506,PhysRevD.100.043504}, such that it is rewritten as
\begin{align}
    R=\frac{P(A,B)}{P(A)P(B)}=\frac{P(A|B)}{P(A)}=\frac{P(B|A)}{P(B)},\label{eq:R2}
\end{align}
where the probabilities of $A$ and $B$ are conditioned on the same underlying model. Noting that $R$ is invariant to the order of $A$ and $B$, Equation \ref{eq:R2} explicitly renders $R\!\gg\!1$ as expressing higher confidence in dataset $A$ in the presence of dataset $B$. Conversely, $R\!\ll\!1$ indicates lower confidence in $A$ in the presence of $B$. Moreover, the value of $R$ has previously been interpreted using the Jeffreys's scale \citep[see, e.g.][]{PhysRevD.98.043526,PhysRevD.100.043504,10.1093/mnras/stab1670}. However, \citet{PhysRevD.100.043504} outline that $R$ is dependent on the choice of priors, indicating a serious limitation of interpreting it on a static scale. Combining Equations \ref{eq:bayes-theorem} and \ref{eq:R2} results in
\begin{align}
    R=\int\frac{\mathcal{P}_A(\theta)\mathcal{P}_B(\theta)}{\pi(\theta)}~d\theta=\left<\frac{\mathcal{P}_A}{\pi}\right>_{\mathcal{P}_B}=\left<\frac{\mathcal{P}_B}{\pi}\right>_{\mathcal{P}_A},
\end{align}
which explicitly shows the dependence of $R$ on $\pi$. Specifically, $R$ can be increased by extending the widths of the priors, reducing the observed tension. Therefore, $R$ can indicate a false positive, where two datasets are incorrectly observed \orange{to agree}. Conversely, false negatives are not possible, since the prior dependency only increases $R$. Hence, if $R$ indicates that two datasets are in tension \red{then} it has captured discordance \red{as long as the prior does not impinge upon the posterior bulk}. Since narrow priors decrease the value of $R$, a practitioner could find a definitive lower bound by iteratively applying the narrowest priors that do not significantly \red{alter} the shape of the posterior. However, such an approach uses a prior that depends on the posterior, which, as a Bayesian practitioner, is not desirable.


\subsection{Bayesian Suspiciousness}\label{sec:suspiciousness}
In response to the limitations of $R$, \citet{PhysRevD.100.043504} introduced a method that divides it in two parts, one that captures the unlikeliness of the dataset posteriors matching given the prior widths, and another that captures the actual mismatch between the dataset posteriors. The first part is the information ratio
\begin{align}
    \log I=\mathscr{D}_A+\mathscr{D}_B-\mathscr{D}_{AB},~\text{where}
\end{align}
\begin{align}
    \mathscr{D}_D=\int\mathcal{P}_D(\theta)\log \frac{\mathcal{P}_D(\theta)}{\pi(\theta)}~d\theta=\left< \log\frac{\mathcal{P}_D}{\pi} \right>_{\mathcal{P}_D}\label{eq:kl-divergence}
\end{align}
is the Kullback-Leibler Divergence \citep{10.1214/aoms/1177729694}. Given some dataset $D$, $\mathscr{D}_D$ quantifies the information gained due to the compression of $\mathcal{P}$ compared to $\pi$ (the average amount of information provided by the posterior). Since $I$ increases similarly to $R$ when the widths of the priors are extended, dividing $I$ from $R$ eliminates the prior dependency issue. The resulting value is the Bayesian suspiciousness
\begin{align}
    \log S=\log R - \log I,\label{eq:bayesian-suspiciousness}
\end{align}
which quantifies the actual mismatch between the dataset posteriors. $S$ can be interpreted as the version of $R$ that corresponds to the narrowest priors that do not significantly impinge on the posterior bulk \citep{{PhysRevD.100.043504,10.1093/mnras/stab1670}}. Therefore, $S\!\ll\!1$ indicates discordance and $S\!\gg\!1$ indicates suspicious concordance.

In the case of Gaussian posteriors, the quantity $d-2\log S$ follows a $\chi^2_d$ distribution, where $d$ is the effective number of parameters (dimensions) constrained by both datasets. If the posteriors are non-Gaussian, the chi-squared interpretation is no longer exact. Assuming that the posteriors are approximately Gaussian, we can compute the probability of observing the measured suspiciousness under the assumption that the datasets are concordant. For the $\chi^2_d$ distribution, we compute the combined dimensionality
\begin{align}
    d=\tilde{d}_A+\tilde{d}_B-\tilde{d}_{AB},~\text{where}
\end{align}
\begin{align}
    \tilde{d}_D=2\left<\left(\log\frac{\mathcal{P}_D}{\pi}\right)^2\right>_{\mathcal{P}_D}-~~~2\left<\log\frac{\mathcal{P}_D}{\pi}\right>^2_{\mathcal{P}_D}\label{eq:bayesian-dimensionality}
\end{align}
is the Bayesian model dimensionality \citep{PhysRevD.100.023512}. We then assign the $p$-value of the $\chi^2_d$ distribution as
\begin{align}
    p_T&=\int^\infty_{d-2\log S}\chi^2_d(x)~dx\\
    &=\int^\infty_{d-2\log S}\frac{x^{d/2-1}e^{-x/2}}{2^{d/2}\Gamma(d/2)}~dx=\frac{\Gamma\left(\frac{d}{2},\frac{d-2\log S}{2}\right)}{\Gamma(d/2)},\label{eq:bayesian-tension-probability}
\end{align}
where $\Gamma(s)$ and $\Gamma(s,x)$ are the regular and upper incomplete gamma functions, respectively. To communicate the significance of the observation, we use the simple Gaussian conversion \citep{10.1093/mnras/stab1670}
\begin{align}
    N_\sigma=\sqrt{2}~\text{Erf}^{-1}(1-p_T),\label{eq:p-to-sigma}
\end{align}
where $\text{Erf}^{-1}$ is the inverse error function. Given a $p$-value from a standard 1D Gaussian distribution that \orange{lies} on the edge of the 95.45 percent confidence region ($2\sigma$), Equation \ref{eq:p-to-sigma} returns the corresponding tension $N_\sigma=2$.

We note that $N_\sigma$ is distinct from the saddle point between 1D or 2D marginalised posteriors (such as those in corner plots). Such visualisations, which one often finds intuitive, can hide existing tension, whereby posteriors that agree in some parameterisation can instead disagree after a parameter transformation \citep[see, e.g. Figure 7 in][]{PhysRevD.100.043504}. Whilst visually intuiting this becomes difficult in higher dimensions, Bayesian suspiciousness is invariant to the choice of parameterisation \citep{PhysRevD.100.043504} \orange{and is therefore} an effective tool for assessing the tension between datasets.

\section{Samples}\label{sec:nvss-and-racs}
Before analysis, we bin all survey samples into sky pixels of equal area using \textsc{Healpy} \citep{2005ApJ...622..759G,Zonca2019}, a \textsc{Python} implementation of \textsc{Healpix}\footnote{\url{https://healpix.sf.net}}. We use $N_\text{side}\!=\!64$, which corresponds to 49\,152 healpixels, each with an area of about 0.83~deg$^2$.

\begin{figure*}
    \includegraphics[trim=0 0 0 55, clip, width=2\columnwidth]{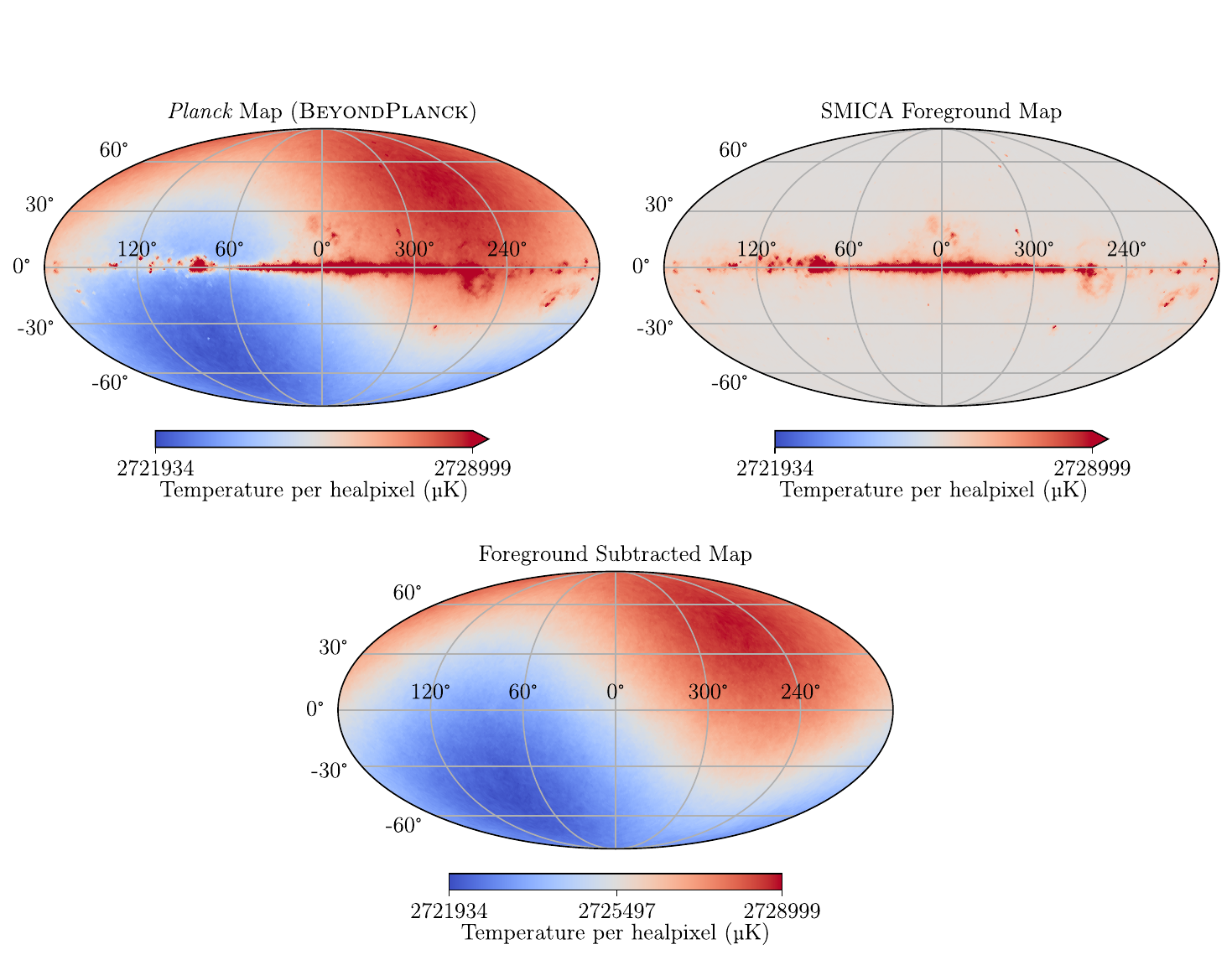}
    \caption{Mollweide sky projections of the \textit{Planck} component maps in Galactic coordinates, binned into 49\,152 healpixels with $N_\text{side}\!=\!64$. \textit{Left:} the \textsc{BeyondPlanck} Data Release II temperature map from the LFI 30~GHz \textit{Planck} observations, containing the dipole and foreground galaxy component. \textit{Right:} the \textsc{SMICA} separated foreground component map from the \textit{Planck} PR3 data release. \textit{Bottom:} the \textsc{BeyondPlanck} LFI~30 GHz temperature map after subtracting the \textsc{SMICA} foreground component, where the middle tick of the colour bar is the mean temperature of the map.}\label{fig:cmb-maps}
\end{figure*}

\subsection{The \textit{Planck} LFI Cosmic Microwave Background}\label{sec:planck-intro}
The European Space Agency's \textit{Planck}\footnote{\url{https://www.esa.int/Planck}} mission observed the CMB at bands centred at 30, 44 and 70~GHz using the Low Frequency Instrument \citep[LFI;][]{2010A&A...520A...4B,2011A&A...536A...3M} and 100, 143, 217, 353, 545, and 857~GHz using the High Frequency Instrument \citep[HFI;][]{2010A&A...520A...9L,2011A&A...536A...4P}. The mission was conducted between 2009 and 2013 using the \textit{Planck} satellite \citep{2010A&A...520A...1T,2011A&A...536A...1P}, producing the 2013, 2015 and 2018 data releases covering the entire sky \citep{2014A&A...571A...1P,2016A&A...594A...1P,2020A&A...641A...1P}.

The first all-sky survey of the CMB was conducted by the Cosmic Background Explorer \citep[COBE;][]{1982OptEn..21..769M,1992ApJ...397..420B} satellite using the Differential Microwave Radiometers \citep[DMR;][]{1990ApJ...360..685S,1992ApJ...391..466B} instrument, and measured our peculiar motion derived from the CMB dipole as $371\pm1$~km~$\text{s}^{-1}$ towards $(l,b)=(264.14\pm0.15^\circ,48.26\pm0.15^\circ)$ \citep{1996ApJ...473..576F}. Subsequently, the Wilkinson Microwave Anisotropy Probe \citep[WMAP;][]{2003ApJ...583....1B} satellite constrained this motion to $369.0\pm0.9$~km~$\text{s}^{-1}$ towards $(l,b)\!=\!(263.99\pm0.14^\circ, 48.26\pm0.03^\circ)$ \citep{2009ApJS..180..225H}. The final \textit{Planck} 2018 data release produced the current highest precision measurement of our peculiar motion derived from the CMB dipole of $v_\text{CMB}\!=\!369.82\pm0.11$~km~$\text{s}^{-1}$ towards $(l,b)\!=\!(264.021\pm0.011^\circ, 48.253\pm0.005^\circ)$ \citep{2020A&A...641A...1P,2020A&A...641A...3P}. Whilst these measurements of the CMB dipole are consistent with each other, they are in general disagreement with the measured amplitude of the dipole in radio and infrared surveys \citep[\newred{for a recent review see, e.g.}][]{PEEBLES2022169159,2022JHEAp..34...49A,KumarAluri_2023,2025NatRP...7...68S}.

\red{To compute the Bayesian evidence ratio and Bayesian suspiciousness used to quantify tension, we must re-analyse the CMB.} We use the \textit{Planck} LFI observations of the CMB as presented by the \textsc{BeyondPlanck}\footnote{\url{https://beyondplanck.science}} Data Release II \citep{2023A&A...675A...1B}. These CMB maps have had the mean temperature subtracted and contain the dipole and foreground galaxy. To separate these components, the \textit{Planck} Collaboration applied four component separation algorithms, \textsc{SMICA} \citep{2008ISTSP...2..735C}, \textsc{Commander} \citep{Eriksen_2004,Eriksen_2008}, \textsc{NILC} \citep{10.1111/j.1365-2966.2011.19770.x,2013MNRAS.435...18B} and \textsc{SEVEM} \citep{Leach_2008,Fern_ndez_Cobos_2012}, to construct foreground intensity maps as part of their Public Data Release~3 \citep[PR3;][]{2020ipac.data.I558P}. We use these to produce 12 \textit{Planck} temperature maps of the CMB at LFI frequencies 30, 44 and 70~GHz by subtracting the foreground intensity component according to each algorithm. Here, we add the mean CMB temperature $\overline{T}_\text{CMB}\!=\!2.7255$~K to each healpixel to account for unphysical negative temperatures. We find that the choice of frequency and component separation algorithm has no significant impact on the results. Therefore, we focus on the \textit{Planck} LFI temperature map at 30~GHz subtracted using the \textsc{SMICA} algorithm (see Figure \ref{fig:cmb-maps}).

\begin{figure*}
	\includegraphics[trim=0 0 0 70, clip, width=2\columnwidth]{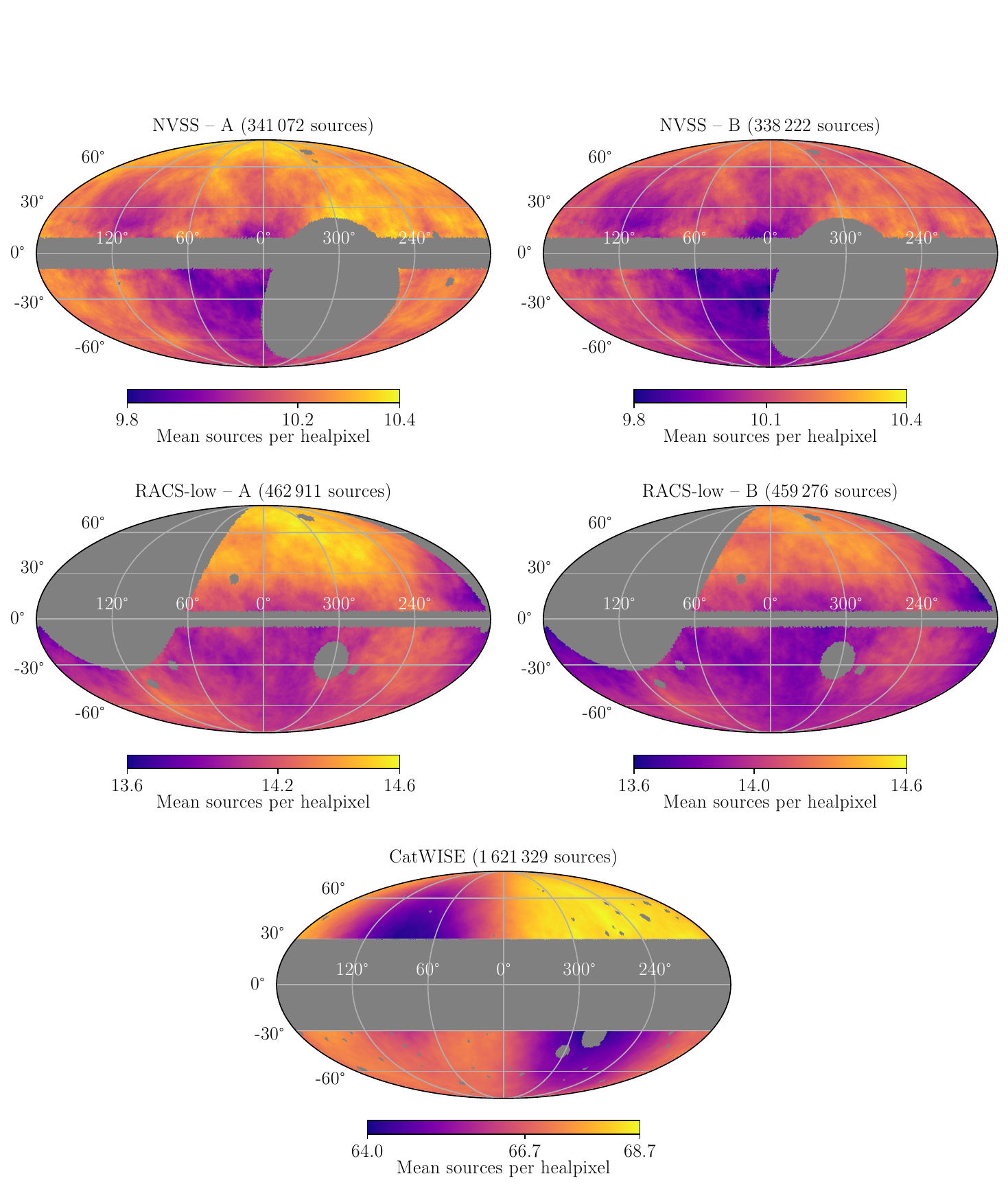}
    \caption{Smoothed Mollweide sky projections of NVSS, RACS-low and CatWISE in Galactic coordinates, binned into 49\,152 healpixels with $N_\text{side}\!=\!64$. The value at each healpixel is the mean number of sources within one steradian. The middle tick of each colour bar is the mean number of sources of the map, \orange{before} smoothing. \textit{Top:} the A and B variants of NVSS. \textit{Middle:} the A and B variants of RACS-low. \textit{Bottom:} CatWISE.}\label{fig:nvss-racs-maps}
\end{figure*}

\subsection{The NRAO VLA Sky Survey}\label{sec:nvss-intro}
NVSS is a 1.4~GHz continuum survey of 1.8 million radio sources, most of which are radio galaxies \citep{Condon_1998}. The survey was conducted between 1993 and 1997 using the National Radio Astronomy Observatory Very Large Array (NRAO VLA), covering the entire sky north of declination $\delta\!\geq\!-40^{\circ}$.

The first measurement of the cosmic dipole using NVSS was done by \citet{2002Natur.416..150B}, where they found a dipole with a direction that agrees with the CMB expectation, but an amplitude that is about 1.5--2 times greater. Subsequently, NVSS has been widely used to extract measurements of the cosmic dipole, with directions that generally agree with that of the CMB, but amplitudes that are over 2 times greater \citep[see, e.g.][]{Singal_2011,10.1111/j.1365-2966.2012.22032.x,Rubart_2013,10.1093/mnras/stu2535,Tiwari_2015,10.1093/mnras/stx1631}. This discrepancy \orange{is greater in recent works, where the amplitude of the NVSS dipole is} approximately 3 times the kinematic expectation \citep[][\citeo]{Secrest_2022,Wagenveld_2023}.

For this study, we use the NVSS samples as presented by \citeo~(see Figure \ref{fig:nvss-racs-maps}). The `A' variant of NVSS of 341\,072 sources is obtained from a flux density distribution of $15~\text{mJy}\!\leq\!S\!\leq\!1000~\text{mJy}$. The lower limit reduces the deviation of the average source density by declination from the mean source density \citep[in line with, e.g.][]{Tiwari_2016,Wagenveld_2023} and the upper limit excludes bright, extended radio sources \citep[see, e.g.][]{10.1046/j.1365-8711.2002.05163.x,Cheng_2024}. Masking is applied to: the Galactic plane ($|b|\!\leq\!10^{\circ}$); an additional degree above the survey declination limit ($-40^{\circ}\!\leq\!\delta\!\leq\!-41^{\circ}$); and localised regions of substantially higher source counts, \orange{several} of which correspond to known bright radio sources (for a complete list, see Table 1 in \citeo). The `B' variant of 338\,222 sources is obtained as described above, with the addition of removing relatively local ($z\!<\!0.1$) radio sources by cross-matching with the source catalogue from the Two Micron All Sky Survey Redshift Survey \citep[2MRS;][]{Huchra_2012} and NASA/IPAC Extragalactic Database (NED)\footnote{\url{https://ned.ipac.caltech.edu}}.

\subsection{The Rapid ASKAP Continuum Survey}
RACS is a large-area survey \citep{McConnell_2020}, with the first data-release "RACS-low1" (hereafter, RACS-low) centred at 887.5~MHz with a bandwidth of 288~MHz \citep{Hale_2021}. The RACS-low observations took place between 2019 and 2020 using the Australian Square Kilometre Array Pathfinder (ASKAP). The source catalogue consists of about 2.1 million radio sources in the regions $-80^\circ\!\leq\!\delta\!\leq\!30^\circ$ and $|b|\!>\!5^\circ$, where the data has been convolved to a common angular resolution of 25 arcsec.

An initial measurement of the cosmic dipole by \citet{Darling_2022} used RACS-low with the Very Large Array Sky Survey \citep[VLASS;][]{Lacy_2020} by combining them at $\delta\!=\!0^{\circ}$ and scaling the RACS-low fluxes at 887.5~MHz to the VLASS fluxes at 3~GHz. With their approach, they found a dipole that was consistent with the CMB in both direction and amplitude. While this result is at odds with the many findings of an anomalously large amplitude \citep[\newred{for a recent review see, e.g.}][]{PEEBLES2022169159,2022JHEAp..34...49A,KumarAluri_2023,2025NatRP...7...68S}, \red{the combined analysis involves two datasets that are not consistent themselves. \citet{Secrest_2022} note that when RACS-low is analysed individually (not in conjunction with VLASS), the dipole amplitude exceeds the kinematic expectation, and that when VLASS is analysed individually, the recovered dipole points towards the south equatorial pole, indicating a systematic effect within the \orange{VLASS} catalogue.} Subsequent measurements of the RACS-low dipole are in line with the general, anomalous trend, finding dipoles with a direction in agreement with the CMB expectation but an amplitude that is three or more times greater \citep[see, e.g.][\citeo]{10.1093/mnras/stad2161,10.1093/mnras/stae414,Wagenveld_2023}.

In line with Section \ref{sec:nvss-intro}, we use the RACS-low samples as presented by \citeo~(see Figure \ref{fig:nvss-racs-maps}). The `A' variant of RACS-low of 462\,911 sources from the flux density distribution $15 \text{ mJy}\!\leq\!S\!\leq\!1000 \text{ mJy}$ accounts for source density declination dependence and excludes bright radio sources. Masking is applied to: partially masked pixels along the edge of the Galactic plane mask ($|b|\!\leq\!5^\circ$); a disc of radius $13^\circ$ centred on the southern pole; an additional degree above the survey declination limit ($-77^{\circ}\!\leq\!\delta\!\leq\!-41^{\circ}$); and several localised regions with substantially high source counts (see Table 1 in \citeo). The `B' variant of 459\,276 sources is obtained as described above, with the addition of removing relatively local radio sources as per the rationale described in Section \ref{sec:nvss-intro}.

\subsection{The CatWISE Quasar Sample}
The CatWISE2020 catalogue \citep[hereafter, CatWISE;][]{Eisenhardt_2020} of 1.9~billion sources was derived from all-sky observations conducted on NASA's WISE spacecraft \citep{Wright_2010}. The data was collected between 2010 and 2018 at 3.4 and 4.6~\textmu m (W1 and W2) as part of the WISE and Near-Earth Object WISE \citep[NEOWISE;][]{Mainzer_2014} surveys.

The first measurement of the cosmic dipole with CatWISE selected 1\,355\,352 quasars from a magnitude cut of $9\!<\!\text{W1}\!<\!16.4$ using the colour criterion $\text{W1}-\text{W2}\!>\!0.8$ and a catalogue mask \citep{Secrest_2021}. They found a dipole amplitude over twice as large as expected, rejecting the kinematic interpretation of the CMB dipole with a significance of $4.9\sigma$. In a subsequent study \red{using a deeper CatWISE sample and a more conservative test statistic,} \citet{Secrest_2022} found a dipole exceeding the kinematic expectation by a factor of about two with a significance of $4.4\sigma$. This result was corroborated by the Bayesian analysis of a similar CatWISE quasar sample in \citet{10.1093/mnras/stad2322}, who found a dipole 2.7 times the expectation with a significance of $5.7\sigma$. \red{Whilst the existence of extraneous power was not shown, \citet{Abghari_2024}} \green{suggest} \red{that the ecliptic latitude bias in the CatWISE quasar sample implies the presence of comparable power in other multipoles, and that if this power exists the significance of the aforementioned results would be reduced.}

We employ the CatWISE quasar sample as presented by \citet{Secrest_2022}, \orange{before} their cross-matching with NVSS and corrections for ecliptic bias (see Figure \ref{fig:nvss-racs-maps}). The sample of 1\,621\,329 quasars is obtained from the magnitude cut of $9\!<\!\text{W1}\!<\!16.5$ (equivalent to $S\!>\!0.078$~mJy) using the colour criterion $\text{W1}\!-\!\text{W2}\!>\!0.8$ from \citet{Stern_2012}. Masking is applied to: the Galactic plane ($|b|\!<\!30^\circ$); known bright sources; and areas with image artefacts or poor-quality photometry. Here, we note that the mask from \citet{Secrest_2022} is a minor revision to that from \citet{Secrest_2021}. Our cross-matching approach from Section \ref{sec:nvss-intro} selects 297 low-redshift sources for removal, which is less than 0.02 percent of the total sources. Since we find that removing these sources has no significant effect on the results, we focus on the sample before cross-matching.

\section{Approach}\label{sec:nested-sampling-approach}
We require a nested sampling approach to analyse the cosmic dipole in \textit{Planck}, NVSS, RACS-low and CatWISE. For Bayesian tensions, we also require machinery that couples the dipole parameters between the datasets during joint analysis. We outline our approach below.

\subsection{Parameter Spans}
Fitting a dipole amplitude $\mathcal{D}$, dipole direction $(l,b)$ and mean source count $\bar{N}$ for a single survey using nested sampling requires a span of free parameters of the form $\Theta_1\!=\!\{\mathcal{D},l,b,\bar{N}\}$. However, when analysing data from multiple surveys, additional parameters are required. Radio surveys are conducted under variable observation conditions, including proximity to the sun and temperature changes throughout the day, and with different sky coverage, centre frequencies, wavebands, observation depths and beam configurations. Despite these observing differences, \red{under the CP,} the underlying population of objects, at a given frequency and sensitivity, is the same. None the less, each survey produces a unique \textit{catalogue} of these sources with mean source count $\bar{N}$ and flux density spectra, with distinct $x$ and $\alpha$. \orange{Since} the dipole amplitude \orange{is dependent} on $x$ and $\alpha$ \orange{(recall from Equation~\ref{eq:dipole-amplitude})}, each survey will have a different \red{expected kinematic amplitude}.

In line with this, previous studies conducting a joint Bayesian analysis of the dipole of two datasets have used a parameter span of the form $\Theta_*\!=\!\{\mathcal{D}_1,\mathcal{D}_2,l,b,\bar{N}_1,\bar{N}_2\}$, where the subscript\orange{s denote} the \orange{parameters} for each survey. Whilst the span $\Theta_*$ describes the unique dipole amplitude of \orange{each} survey and jointly fits the direction, fitting \orange{the} dipole amplitude\orange{s separately} is not compatible with Bayesian tensions. \orange{To jointly fit the dipole amplitudes, we parametrise them using our heliocentric motion \citep[see also][who used a similar approach]{wagenveld2025}.} Assuming an observer traveling at $v_\text{CMB}$ with $\beta_\text{CMB}\!=\!v_\text{CMB}/c$, a flux-limited catalogue $k$ with intrinsic $x_k$ and $\alpha_k$ has an expected kinematic amplitude of
\begin{align}
\tilde{\mathcal{D}}_k=(2+x_k[1+\alpha_k])\beta_\text{CMB}.
\end{align}
To find the dipole amplitude of the survey at some other velocity $v$, we linearly scale the expected \orange{kinematic} amplitude, \orange{such that}
\begin{align}
\mathcal{D}_k(v)=\tilde{\mathcal{D}}_k\cdot v/v_\text{CMB}.
\end{align}
Our parameter span $\Theta_*$ becomes $\{\mathcal{D}_1(v),\mathcal{D}_2(v),l,b,\bar{N}_1,\bar{N}_2\}$ \orange{and} simplifies to $\Theta_2\!=\!\{v,l,b,\bar{N}_1,\bar{N}_2\}$ due to the shared dependence on $v$, \orange{which} satisfies the physical and Bayesian tension requirements. \orange{Here, the mean source count parameters $\bar{N}_1$ and $\bar{N}_2$ are dataset-specific nuisance parameters that do not contribute to the tension.} In \orange{this work}, we use the parameter span $\Theta_1$ for the individual analyses and $\Theta_2$ for the joint analyses. Additionally, for the CatWISE quasar sample, we include the ecliptic bias $\Upsilon_\text{ecl}$ nuisance parameter \citep[see, e.g.][]{Secrest_2021,10.1093/mnras/stad2322}, which does not contribute to the tension. We determine the expected \orange{kinematic} amplitude $\tilde{\mathcal{D}}_k$ of each survey a priori, which we explain below.

\subsection{Expected Dipole Amplitude}\label{sec:expected-amp}
Since the amplitude of the CMB dipole is $v_\text{CMB}/c$ \citep{PhysRev.174.2168}, the expected amplitude for \textit{Planck} is simply $\tilde{\mathcal{D}}_\textit{Planck}\!=\!\beta_\text{CMB}$. For our samples of NVSS, RACS-low and CatWISE, we use the approach from Section 4.1.2 of \citeo~\citep[see also,][]{10.1093/mnras/stad3706} to compute the expected amplitude:
\begin{enumerate}
    \renewcommand{\labelenumi}{\roman{enumi}.}
    \item For each source with flux density $S_i$ and error $\sigma_{S_i}$, we regenerate the catalogue to account for uncertainties by drawing a new flux density $S_i^*\!\in\!\mathcal{G}(\mu\!=\!S_i,\sigma\!=\!\sigma_{S_i})$;
    \item We compute the number of regenerated sources $N_i$ above some limiting flux density $S_0$;
    \item We Doppler boost each source by $S_i^* \delta^{1+\alpha_i}$ with $\delta\!=\!\gamma_\text{CMB}(1+\beta_\text{CMB})$ where $\alpha_i$ is the measured spectral index. Where we do not have measurements of $\alpha_i$, for example, in NVSS and RACS-low, we instead draw $\alpha_i$ from $\mathcal{G}(\mu\!=\!0.75,\sigma\!=\!0.5)$; and
    \item We compute the number of boosted sources $N_b$ above the same limiting flux density $S_0$, and find the expected amplitude
    \begin{align}
    \tilde{\mathcal{D}}_k=\frac{N_b\delta^2-N_i}{N_i},
    \end{align}
    where $k$ is the survey of interest and the $\delta^2$ term accounts for relativistic aberration \citep{1984MNRAS.206..377E}.
\end{enumerate}
We repeat steps i--iv 100 times with $S_0$ between 15 and 35~mJy for NVSS and RACS-low, and between 0.078 and 0.09~mJy for CatWISE, and compute the mean expected amplitude at each $S_0$. Since there are no sources below the lower flux density limit of the sample, when $S_0$ approaches that limit, the expected amplitude is underestimated due to the boundary effect from regenerating sources. Therefore, we extrapolate the expected amplitude at that limit by computing the gradient of $\tilde{\mathcal{D}}_k$ vs $S_0$. We use a least-squares linear fit with $S_0\!>\!17$~mJy for NVSS, $S_0\!>\!20$~mJy for RACS-low, and $S_0\!>\!0.079$~mJy for CatWISE, arriving at $\tilde{\mathcal{D}}_\text{NVSS}\!=\!4.31\!\times\!10^{-3}$, $\tilde{\mathcal{D}}_\text{RACS-low}\!=\!4.27\!\times\!10^{-3}$ and $\tilde{\mathcal{D}}_\text{CatWISE}\!=\!7.25\!\times\!10^{-3}$. Here, the CatWISE amplitude is in line with that from \citet{Secrest_2022}, and the NVSS and RACS-low amplitudes are identical to those from \citeo. Since there is \orange{a} negligible difference between the expected amplitudes of the A and B variants, we use the same value for both.

\subsection{Likelihood Functions}
The likelihood is the \orange{probability} of \orange{observing} the data given a set of parameters, conditioned on an underlying model. In practice, this is the multiplied probabilities of observing each pixel based on an \textit{expected} observation. The expectation value at each pixel $\mathbf{\hat{n}}_i$ is given to first order by the sum of the \red{dipolar modulation}
\begin{align}
\lambda_i=\lambda(\mathbf{\hat{n}}_i)&=\bar{N}(1+\mathbf{d}\cdot\mathbf{\hat{n}}_i)\label{eq:expectation-value-vector}\\
&=\bar{N}(1+\mathcal{D}\cos\theta_i),\label{eq:expectation-value}
\end{align}
where $\theta_i$ is the angle between the dipole and pixel $\mathbf{\hat{n}}_i$.

In line with \citet{10.1093/mnras/stad2322}, we include the ecliptic bias when fitting CatWISE. We multiply Equation \ref{eq:expectation-value} by
\begin{align}
f_\text{ecl}(\hat{\mathbf{n}}_i)\equiv1-\Upsilon_\text{ecl}c_\text{ecl}|b_\text{ecl}(\hat{\mathbf{n}}_i)|,
\end{align}
where the slope $c_\text{ecl}\!=\!9.15\times10^{-4}$ is from \citet{Secrest_2022} and $b_\text{ecl}(\hat{\mathbf{n}}_i)$ is the ecliptic latitude at pixel $\mathbf{\hat{n}}_i$. The independent discretised counts in each pixel of NVSS, RACS-low and CatWISE motivate a Poissonian likelihood, whereby the probability of observing $N_i$ discrete counts in pixel $\mathbf{\hat{n}}_i$ is
\begin{align}
P_\text{discrete}(N_i|\Theta)=\frac{\lambda_i^{N_i}e^{-\lambda_i}}{N_i!},\label{eq:nvss-racs-pixel-prob}
\end{align}
given some set of free parameters $\Theta$. Likewise, the continuous temperature measurements in each pixel of \textit{Planck} motivate a Gaussian likelihood, such that the probability of observing the temperature $N_i$ in pixel $\mathbf{\hat{n}}_i$ is given by
\begin{align}
P_\text{continuous}(N_i|\Theta)= \mathcal{G}_\text{PDF}(x=N_i,\mu=\lambda_i,\sigma=\sigma_D),\label{eq:cmb-pixel-prob}
\end{align}
where $\sigma_D$ is the standard deviation of the data $D$ and $\mathcal{G}_\text{PDF}$ is the Gaussian probability density function. Finally, the likelihood of observing the data, given the parameters $\Theta$, is the product of the probability of observing each unmasked pixel across the set of all unmasked pixels $n_\text{pix}$, such that
\begin{align}
\mathcal{L}(D|\Theta)=\prod_{i=1}^{n_\text{pix}}P(N_i|\Theta).
\end{align}

\subsection{Prior Likelihood Functions}
During nested sampling, the priors $\pi$ for each parameter are constructed given some transformation of a uniform random variable $u\in[0,1]$. Since \orange{we quantify tension using Bayesian suspiciousness,} we opt for wide priors that \orange{do} not impinge upon the posterior bulk:
\begin{align*}
\pi(v)&\sim 20\cdot v_\text{CMB}\cdot u\\
\pi(l)&\sim360\cdot u\\
\pi(b)&\sim\sin^{-1}(2\cdot u-1)\\
\pi(\bar{N})&\sim \bar{N}_*(0.2\cdot u+0.9)\\
\pi(\Upsilon_\text{ecl})&\sim4\cdot u-2.
\end{align*}
\orange{We sample} up to 20 times \orange{$v_\text{CMB}$ to enclose} previous measurements of the dipole \citep[\newred{for a recent review see, e.g.}][]{PEEBLES2022169159,2022JHEAp..34...49A,KumarAluri_2023,2025NatRP...7...68S}. \orange{Moreover,} we choose $\pi(b)$ to give uniform sampling over the sphere. \orange{We find that} a variance of ten percent from \orange{the mean source count of the input data ($\bar{N}_*$) is adequate, as is allowing} up to two times $|\Upsilon_\text{ecl}|$.


We use the above prior and likelihood functions to compute the nested sampling chains, posterior distributions, and Bayesian evidence with the nested sampling Monte Carlo algorithm \textsc{MLFriends} \citep{2016S&C....26..383B,2019PASP..131j8005B} using \textsc{UltraNest}\footnote{\url{https://johannesbuchner.github.io/UltraNest}} \citep{2021JOSS....6.3001B}. Finally, we calculate $\mathcal{Z}$, $\mathscr{D}$ and $\tilde{d}$ using \textsc{Anesthetic}\footnote{\url{https://github.com/handley-lab/anesthetic}} \citep{anesthetic}.

\section{Results}\label{sec:results}
\begin{figure*}
    \centering
        \includegraphics[width=\columnwidth]{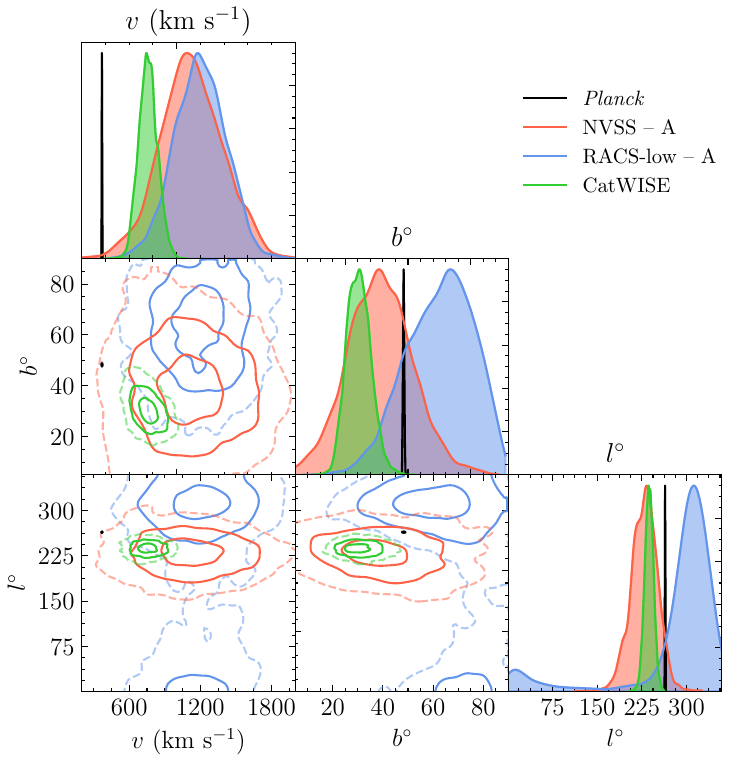}
    \hfill
        \includegraphics[width=\columnwidth]{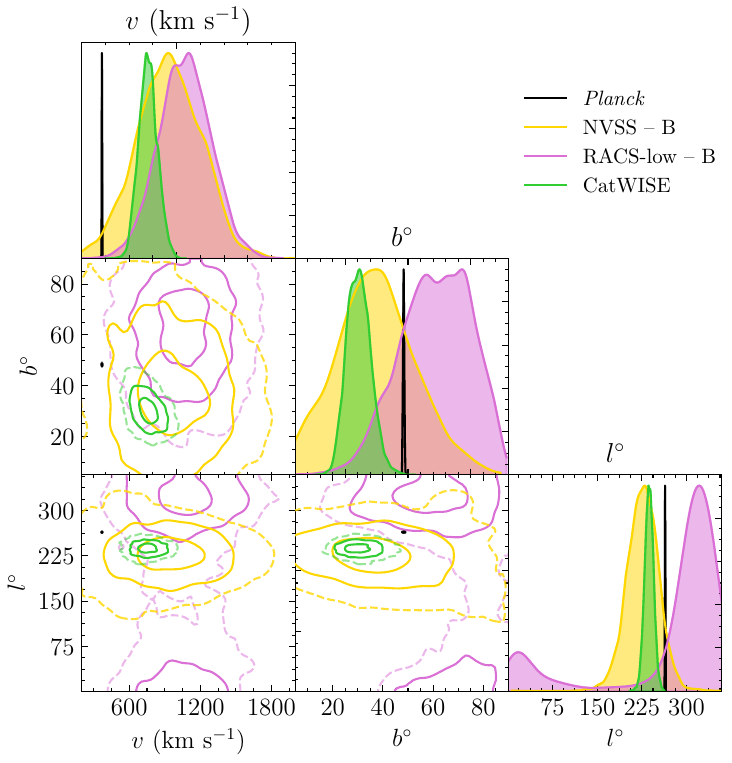}
    \caption{Posteriors from the individual analyses of \textit{Planck}, NVSS, RACS-low and CatWISE. The nuisance parameters are omitted for clarity. The contours of the 2D marginal posteriors are in intervals of $1\sigma$, encompassing 39.4, 86.4 and 98.9 (dashed) percent of the probability distribution. These contours have been smoothed by three percent using a Gaussian kernel. \textit{Left:} the posteriors of \textit{Planck}, CatWISE and the A variants of NVSS and RACS-low. \textit{Right:} the posteriors of \textit{Planck}, CatWISE and the B variants of NVSS and RACS-low.}
    \label{fig:corner-plots}
\end{figure*}

We present the results from each individual and joint analysis of \textit{Planck}, NVSS, RACS-low and CatWISE. Here, we present the inferred dipole and nuisance parameters of the individual analyses in Table \ref{table:results-individual} and the joint analyses in Table \ref{table:results-joint}, including their Bayesian tension statistics. In addition, we visualise the posteriors of the individual analyses in Figure~\ref{fig:corner-plots}.

\begin{table}
\centering
\begin{tabular}{@{\hskip 1pt}l@{\hskip 4.5pt}c@{\hskip 4.5pt}c@{\hskip 4.5pt}c@{\hskip 4.5pt}c@{\hskip 4.5pt}c@{\hskip 4.5pt}c@{\hskip 4.5pt}c@{\hskip 1pt}}
\hline \\ [-3ex]
 &&\multicolumn{3}{c}{Inferred Dipole}&\multicolumn{2}{c}{Nuisance Parameters}\\
 \hline \\ [-3ex]
 Dataset&&$v$ [km $\text{s}^{-1}$]&$l~[^\circ]$&$b~[^\circ]$&$\bar{N}~[\text{px}\substack{-1\\{}}]$&$\Upsilon_\text{ecl}$\\
 \hline \\ [-2.7ex]
 \textit{Planck}&--&$370\!\pm\!3$&${264.1}_{-0.7}^{+0.8}$&${48.2}\!\pm\!{0.5}$&${2725497}_{-16}^{+17}$&--\\[1ex]
 NVSS&A&${1119}_{-520}^{+523}$&${230}_{-45}^{+40}$&${39}_{-27}^{+26}$&${10.16}\!\pm\!{0.04}$&--\\[1ex]
 "&B&${932}_{-542}^{+508}$&${227}_{-55}^{+50}$&${37}_{-31}^{+33}$&${10.07}\!\pm\!{0.04}$&--\\[1ex]
 RACS-low&A&${1194}_{-418}^{+403}$&${307}_{-298}^{+43}$&${64}_{-29}^{+21}$&${14.17}\!\pm\!{0.05}$&--\\[1ex]
 "&B&${1073}_{-439}^{+424}$&${313}_{-307}^{+44}$&${62}_{-30}^{+23}$&${14.06}\!\pm\!{0.05}$&--\\[1ex]
 CatWISE&--&${757}_{-144}^{+161}$&${237}_{-15}^{+14}$&${30}_{-8}^{+10}$&${68.6}\!\pm\!0.2$&${0.99}\!\pm\!{0.07}$\\
 \hline
\end{tabular}
\caption{The inferred dipole and nuisance parameters of the individual analyses of \textit{Planck}, NVSS, RACS-low and CatWISE. The value in each cell is the median of its respective marginalised posterior, with the upper and lower bounds of its $2\sigma$ credible interval. The $\bar{N}$ parameter is in units of counts per healpixel, where the counts are \textmu K for \textit{Planck} and sources for NVSS, RACS-low and CatWISE.}\label{table:results-individual}
\end{table}

\begin{table*}
\centering
\begin{tabular}{@{\hskip 1.5pt}l@{\hskip 5pt}c@{\hskip 5pt}c@{\hskip 5pt}c@{\hskip 5pt}c@{\hskip 5pt}c@{\hskip 5pt}c@{\hskip 5pt}c@{\hskip 5pt}c@{\hskip 5pt}c@{\hskip 5pt}c@{\hskip 5pt}c@{\hskip 5pt}c@{\hskip 1.5pt}}
\hline \\ [-3ex]
 &&\multicolumn{3}{c}{Inferred Dipole}&\multicolumn{3}{c}{Nuisance Parameters}&\multicolumn{5}{c}{Bayesian Statistics}\\
 \hline \\ [-3ex]
 Dataset&&$v$ [km $\text{s}^{-1}$]&$l~[^\circ]$&$b~[^\circ]$&$\bar{N}_1~[\text{px}\substack{-1\\{}}]$&$\bar{N}_2~[\text{px}\substack{-1\\{}}]$&$\Upsilon_\text{ecl}$&$\log{R}$&$\log{I}$&$\log{S}$&$d$&$\sigma$\\
 \hline \\ [-2.7ex]
 \textit{Planck}--NVSS&A&${370}\!\pm\!{3}$&${264.1}\!\pm\!{0.8}$&${48.2}\!\pm\!{0.5}$&${2725498}_{-18}^{+16}$&${10.16}_{-0.03}^{+0.04}$&--&${0.5}\!\pm\!{0.8}$&${4.4}\!\pm\!{0.8}$&${-3.8}\!\pm\!{1.1}$&${2.8}_{-0.5}^{+0.6}$&${2.5}_{-0.5}^{+0.4}$\\[1ex]
 "&B&${370}\!\pm\!{3}$&${264.1}\!\pm\!{0.8}$&${48.2}\!\pm\!{0.5}$&${2725497}\!\pm\!{17}$&${10.08}_{-0.04}^{+0.03}$&--&${2.1}\!\pm\!{0.8}$&${4.1}\!\pm\!{0.8}$&${-2.0}\!\pm\!{1.1}$&${2.9}\!\pm\!{0.6}$&${1.8}\!\pm\!{0.5}$\\[1ex]
 \textit{Planck}--RACS-low&A&${370}\!\pm\!{3}$&${264.1}\!\pm\!{0.8}$&${48.3}\!\pm\!{0.5}$&${2725497}_{-17}^{+18}$&${14.16}\!\pm\!{0.04}$&--&${-3.6}\!\pm\!{0.8}$&${4.4}\!\pm\!{0.8}$&${-8.0}\!\pm\!{1.1}$&${3.1}_{-0.5}^{+0.6}$&${3.6}\!\pm\!{0.4}$\\[1ex]
 "&B&${370}\!\pm\!{3}$&${264.1}\!\pm\!{0.8}$&${48.3}\!\pm\!{0.5}$&${2725498}_{-17}^{+16}$&${14.05}\!\pm\!{0.04}$&--&${-1.1}\!\pm\!{0.8}$&${4.7}\!\pm\!{0.8}$&${-5.8}\!\pm\!{1.1}$&${2.7}_{-0.5}^{+0.6}$&${3.1}\!\pm\!{0.4}$\\[1ex]
 \textit{Planck}--CatWISE&--&${370}\!\pm\!3$&${264.1}\!\pm\!{0.8}$&${48.2}\!\pm\!{0.5}$&${2725498}\!\pm\!{17}$&${68.6}\!\pm\!0.2$&${0.99}_{-0.08}^{+0.07}$&${-7.4}\!\pm\!{0.9}$&${7.4}\!\pm\!{0.8}$&${-14.8}\!\pm\!{1.2}$&${2.8}_{-0.6}^{+0.7}$&${5.1}\!\pm\!{0.3}$\\[1ex]
 NVSS--RACS-low&A&${1063}_{-314}^{+325}$&${268}_{-65}^{+42}$&${61}_{-23}^{+20}$&${10.16}\!\pm\!{0.04}$&${14.17}_{-0.04}^{+0.05}$&--&${0.6}\!\pm\!{0.5}$&${3.7}\!\pm\!{0.5}$&${-3.1}\!\pm\!{0.7}$&${3.0}\!\pm\!{0.6}$&${2.2}\!\pm\!{0.3}$\\[1ex]
 "&B&${909}_{-324}^{+314}$&${270}_{-94}^{+51}$&${62}_{-25}^{+20}$&${10.07}\!\pm\!{0.04}$&${14.06}_{-0.04}^{+0.05}$&--&${0.9}\!\pm\!{0.5}$&${3.6}\!\pm\!{0.5}$&${-2.6}\!\pm\!{0.7}$&${3.4}_{-0.5}^{+0.6}$&${2.0}\!\pm\!{0.3}$\\[1ex]
 NVSS--CatWISE&A&${788}_{-146}^{+151}$&${237}\!\pm\!{14}$&${30}_{-8}^{+10}$&${68.6}\!\pm\!{0.2}$&${10.16}\!\pm\!{0.03}$&${0.99}_{-0.07}^{+0.08}$&${4.0}\!\pm\!{0.6}$&${4.0}\!\pm\!{0.6}$&${0.0}\!\pm\!{0.9}$&${2.8}_{-0.6}^{+0.7}$&${0.9}_{-0.7}^{+0.5}$\\[1ex]
 "&B&${774}_{-151}^{+154}$&${237}_{-15}^{+14}$&${30}_{-8}^{+10}$&${68.6}\!\pm\!{0.2}$&${10.08}\!\pm\!{0.04}$&${0.99}_{-0.08}^{+0.07}$&${4.9}\!\pm\!{0.6}$&${4.0}\!\pm\!{0.6}$&${0.8}\!\pm\!{0.9}$&${3.0}_{-0.6}^{+0.7}$&${0.4}_{-0.3}^{+0.6}$\\[1ex]
 RACS-low--CatWISE&A&${733}\!\pm\!{139}$&${245}_{-16}^{+15}$&${35}_{-9}^{+11}$&${68.6}\!\pm\!{0.2}$&${14.16}\!\pm\!{0.04}$&${0.99}\!\pm\!{0.07}$&${-3.8}\!\pm\!{0.6}$&${5.0}\!\pm\!{0.6}$&${-8.8}\!\pm\!{0.9}$&${3.2}_{-0.6}^{+0.7}$&${3.8}\!\pm\!{0.3}$\\[1ex]
 "&B&${717}_{-143}^{+146}$&${244}_{-17}^{+16}$&${36}_{-9}^{+11}$&${68.6}\!\pm\!{0.2}$&${14.05}\!\pm\!{0.04}$&${0.99}\!\pm\!{0.07}$&${-3.0}\!\pm\!{0.6}$&${4.6}\!\pm\!{0.6}$&${-7.6}\!\pm\!{0.9}$&${2.8}_{-0.6}^{+0.7}$&${3.6}\!\pm\!{0.3}$\\
 \hline
\end{tabular}
\caption{The inferred dipole, nuisance parameters and Bayesian statistics of the joint analyses of \textit{Planck}, NVSS, RACS-low and CatWISE. The value in each cell is the median of its respective marginalised posterior, with the upper and lower bounds of its $2\sigma$ credible interval. The $\bar{N}_1$ and $\bar{N}_2$ parameters are in units of counts per healpixel, where the counts are \textmu K for \textit{Planck} and sources for NVSS, RACS-low and CatWISE.}\label{table:results-joint}
\end{table*}

\subsection{\red{Individual and Joint Analysis of} \textit{Planck}--NVSS}
Within a $2\sigma$ credible interval, we find that the dipole direction of \red{both variants} of NVSS \red{are} consistent with \textit{Planck} (see the first three rows in Table \ref{table:results-individual} and the corresponding contours in Figure \ref{fig:corner-plots}). Despite \red{this} directional agreement, the velocities of both variants are not consistent with \textit{Planck}. The extent of this tension is quantified through the computed Bayesian terms, as described in Section~\ref{sec:nested-sampling-approach} (see the resulting statistics in Table \ref{table:results-joint}). The Bayesian evidence ratios ($\log R$) with their respective information ratios ($\log I$) result in Bayesian suspiciousnesses ($\log S$) of \red{$-3.8\pm1.1$} and \red{$-2.0\pm1.1$} for the A and B variants of \textit{Planck}--NVSS, respectively. Considering the Bayesian dimensionalities ($d$) of \red{${2.8}\substack{+0.6\\-0.5}$} and \red{$2.9\pm0.6$}, the probability of observing the measured suspiciousness corresponds to a \textit{Planck}--NVSS tension of \red{$2.5\substack{+0.4\\-0.5}\sigma$} and \red{$1.8\pm0.5\sigma$}, for the A and B variants, respectively. This places NVSS in \textbf{moderate tension} with \textit{Planck}.

\subsection{\red{Individual and Joint Analysis of} \textit{Planck}--RACS-low}
The dipole directions \red{and velocities} of both RACS-low variants are \red{not} consistent with \textit{Planck} (see rows one, four, and five in Table \ref{table:results-individual} and the corresponding contours in Figure \ref{fig:corner-plots}). We measure the A and B variants of \textit{Planck}--RACS-low to have a $\log S$ of \red{$-8.0\pm1.1$} and \red{$-5.8\pm1.1$}, respectively (see the statistics in Table \ref{table:results-joint}). With the $d$ values of \red{${3.1}\substack{+0.6\\-0.5}$} and \red{$2.7\substack{+0.6\\-0.5}$}, the probability of observing the measured suspiciousness of the A and B variants converts to a \textit{Planck}--RACS-low tension of \red{$3.6\pm0.4\sigma$} and \red{$3.1\pm0.5\sigma$}, respectively. This places RACS-low in \textbf{\red{strong} tension} with \textit{Planck}.

\subsection{\red{Individual and Joint Analysis of} \textit{Planck}--CatWISE}
For CatWISE, the velocity and dipole directions are not consistent with \textit{Planck} (see the first and last rows in Table \ref{table:results-individual} and the corresponding contours in Figure \ref{fig:corner-plots}). We measure \textit{Planck}--CatWISE to have a suspiciousness of $-14.8\pm1.2$ with dimensionality ${2.8}\substack{+0.7\\-0.6}$ (see the statistics in Table \ref{table:results-joint}). Therefore, the probability of observing the measured suspiciousness corresponds to a \textit{Planck}--CatWISE tension of $5.1\pm0.3\sigma$. This places CatWISE in \textbf{severe tension} with \textit{Planck}.

\subsection{\red{Individual and Joint Analysis of} NVSS--RACS-low}
The velocities and dipole directions of both NVSS variants are consistent with their respective RACS-low variants (see rows two--four in Table \ref{table:results-individual} and the corresponding contours in Figure \ref{fig:corner-plots}). We measure the A and B variants to have a $\log S$ of \red{$-3.1\pm0.7$} and \red{$-2.6\pm0.7$}, respectively (see the statistics in Table \ref{table:results-joint}). For the A and B variants, considering the $d$ values of \red{${3.0}\pm0.6$} and \red{$3.4\substack{+0.6\\-0.5}$}, the probability of observing the measured suspiciousness converts to a respective NVSS--RACS-low tension of \red{$2.2\pm0.3\sigma$} and \red{$2.0\pm0.3\sigma$}. This indicates that NVSS and RACS-low are in \textbf{moderate tension}.

\subsection{\red{Individual and Joint Analysis of} NVSS--CatWISE}
Notably, the velocities and dipole directions of both NVSS variants are highly consistent with CatWISE (see rows two, three, and six in Table \ref{table:results-individual} and the corresponding contours in Figure \ref{fig:corner-plots}). We measure the A and B variants of NVSS--CatWISE to have a $\log S$ of \red{$0.0\pm0.9$} and \red{$0.8\pm0.9$}, respectively (see the statistics in Table \ref{table:results-joint}). Considering that $d$ is \red{${2.8}\substack{+0.7\\-0.6}$} and \red{${3.0}\substack{+0.7\\-0.6}$} for the A and B variants, the probability of observing the measured suspiciousness corresponds to a respective NVSS--CatWISE tension of \red{${0.9}\substack{+0.5\\-0.7}\sigma$} and \red{${0.4}\substack{+0.6\\-0.3}\sigma$}. This indicates that there is \textbf{no significant tension} between NVSS and CatWISE.

\subsection{\red{Individual and Joint Analysis of} RACS-low--CatWISE}
Finally, the velocities of both RACS-low variants are consistent with CatWISE; however, their dipole directions are not (see the last three rows in Table \ref{table:results-individual} and the corresponding contours in Figure \ref{fig:corner-plots}). The A and B variants of RACS-low--CatWISE has a suspiciousness of \red{$-8.8\pm0.9$} and \red{$-7.6\pm0.9$}, with dimensionality \red{${3.2}\substack{+0.7\\-0.6}$} and \red{${2.8}\substack{+0.7\\-0.6}$}, respectively (see the statistics in Table \ref{table:results-joint}). Hence, the probability of observing the measured suspiciousness corresponds to a RACS-low--CatWISE tension of \red{$3.8\pm0.3\sigma$} and \red{$3.6\pm0.3\sigma$}, for the A and B variants, respectively. This indicates that RACS-low and CatWISE are in \textbf{strong tension}.

\section{Discussion \& Conclusions}\label{sec:discussion}

\subsection{Tension with \textit{Planck} Compared to Recent Literature}\label{sec:discussion-corroboration}

We compare \orange{our observed} tensions \orange{between} \textit{Planck} \orange{and} NVSS, RACS-low and CatWISE to \orange{their respective} analyses in the literature \orange{that examine them against} the CMB expectation.

The A variants of NVSS and RACS-low are in moderate \orange{and strong} tension with \textit{Planck}, \orange{respectively,} primarily due to their inferred velocities \orange{significantly exceeding} the kinematic expectation. This is in general consensus with previous analyses of NVSS and RACS-low in the literature that find an excessive dipole amplitude. Specifically, the NVSS A variant velocity, of \red{${3.0}\pm{1.4}$} times \orange{$v_\text{CMB}$}, agrees with the amplitudes greater than about two times \citep[][]{Singal_2011,10.1111/j.1365-2966.2012.22032.x,Rubart_2013,10.1093/mnras/stu2535,Tiwari_2015,10.1093/mnras/stx1631} and three times the kinematic expectation \citep[][\citeo]{Secrest_2022,Wagenveld_2023}. Likewise, the RACS-low A variant velocity, of \red{$3.2\pm1.1$} times \orange{$v_\text{CMB}$}, agrees with the recent amplitudes \orange{over} three times the kinematic expectation \citep[see, e.g.][\citeo]{10.1093/mnras/stad2161,10.1093/mnras/stae414,Wagenveld_2023}.

After cross-matching the radio catalogues with local source catalogues and removing \orange{the matches}, we find that the inferred velocity decreases. Specifically, the velocities of the B variants of NVSS and RACS-low drop to \red{$2.5\substack{+1.4\\-1.5}$} and \red{$2.9\substack{+1.1\\-1.2}$} times \orange{$v_\text{CMB}$}; a reduction of about \red{17} and \red{10} percent, respectively. This amplitude reduction after removing local sources is in line with the 10--15 percent reduction observed by \citeo. Since the decrease brought \orange{their velocities} closer to that of \textit{Planck}, we observe a reduction in tension by about \red{$0.7\sigma$ and $0.5\sigma$ for NVSS and RACS, respectively}. Despite the reduction, the tension between the B variants and \textit{Planck} is still \red{considerable}, with the \red{strong} median \textit{Planck}--RACS-low tension of \red{$3.1\sigma$} exceeding the \red{moderate} \textit{Planck}--NVSS tension of \red{$1.8\sigma$}. This difference in tension between NVSS and RACS-low corroborates the model comparison in \citeo. For the B variants, \red{they find that} RACS-low favours a dipole with free parameters rather than a dipole with direction and amplitude fixed to the CMB expectation, and that NVSS favours a dipole fixed to the CMB expectation rather than a free dipole. Here, RACS-low favouring the free dipole whilst NVSS favours the CMB dipole is \orange{in} line with our observations of greater \textit{Planck}--RACS-low tension compared to \textit{Planck}--NVSS. We further explore the difference between the NVSS and RACS-low dipoles in Section~\ref{sec:internal-tensions}.

Turning to CatWISE, the inferred dipole amplitude corresponds to a velocity of $2.0\pm0.4$ times \orange{$v_\text{CMB}$} (see Table \ref{table:results-individual}). \orange{This agrees} with the dipole amplitudes \orange{over} about two times the \orange{kinematic} expectation found in recent analyses of CatWISE in the literature \citep[see, e.g.][]{Secrest_2021,Secrest_2022,10.1093/mnras/stad2322}. In place of a null hypothesis comparison or the probability of exceeding \orange{$v_\text{CMB}$}, we compare our CatWISE dipole to a fit of the \textit{Planck} data. We find that \textit{Planck} and CatWISE are in $5.1\sigma$ tension under the kinematic interpretation, which corroborates the recent \orange{CatWISE dipole} measurements observed at a significance of 4.4--$5.7\sigma$ compared to the kinematic \orange{expectation} \citep[see, e.g.][]{Secrest_2021,Secrest_2022,10.1093/mnras/stad2322}. Considering the corroboration of our analyses with recent literature, the tensions we present here uphold the consensus view that the cosmic dipole anomaly presents a significant challenge to the standard model of cosmology.

\subsection{\red{Effect of Cross-Matched Sources on the Tension with \textit{Planck}}}\label{sec:discussion-redshift}

\red{The \orange{radio and infrared} samples in this work contain some amount of intrinsic} heterogeneous and anisotropic local structure. This structure may impart a dipole in \textit{any} direction with an amplitude unlike that from our heliocentric motion, inducing tension with \red{the \textit{Planck} CMB. Assuming that we are observers in $\Lambda$CDM where the CP holds on large scales, removing local structure should bring the samples more in line with the homogeneous and isotropic background of the distant Universe. This is captured by the reduction of tension between \textit{Planck} and both NVSS and RACS-low after \orange{removing local sources.} However, there remains considerable tension (recall from Section \ref{sec:discussion-corroboration}). Whilst the origin of this outstanding tension is unknown, it appears unlikely that \orange{it can be explained in its entirety by} local sources remaining in the samples, which we explain below.}

\green{Firstly, due to the incompleteness of 2MRS, which shows a rapid decline after $z\approx0.02$, some radio sources \cyan{will} not have an optical counterpart. Therefore, when removing local sources with cross-matching, there may be a population of low redshift sources remaining in the radio sample. Moreover, associating multi-component radio sources with the same optical or near-infrared host galaxy presents an additional challenge when cross-matching.} Depending on the resolution of the radio survey, a galaxy source may be resolved as multiple components owing to spatially extended radio-loud AGN stretching beyond the galaxy centre (radio lobes). These complex-structured sources are difficult to cross-match with optical or near-infrared observations, since only the central component will match the host galaxy. Increasing the cross-matching radius to account for radio lobes involves matching up to three radio components per optical or near-infrared component, varying case-by-case. This intractable process is a substantial unsolved problem for radio surveys and is beyond the scope of this work \citep[see, e.g. recent literature on this issue;][]{2018MNRAS.478.5547A,2019A&A...622A...2W,2024PASA...41...27G}. \green{Considering the incompleteness of 2MRS and the limitations associated with radio to optical or near-infrared cross-matching, it is highly likely that there is a population of low redshift sources remaining in the radio samples that we use in this work.}


\red{In light of the above, can a population of low redshift sources remaining in the samples explain the moderate and strong tensions that we observe? Regarding the radio \orange{samples}, since radio sources \orange{above $\sim\!10$}~mJy are predominantly at moderate \orange{($>\!0.1$)} redshift \citep[see, e.g.][]{1984ApJ...287..461C,2025RSPTA.38340027S}, it appears unlikely that such a population can explain the tensions that we observe with \textit{Planck}.}



\red{\orange{Regarding CatWISE, since their near-infrared components are compact} the sample does not suffer from the same cross-matching limitations as the radio surveys. 
Since the mean redshift of the CatWISE quasars is 1.2 with 99 percent of the quasars having redshift $z\!>\!0.1$ \citep[see Figure 3 in][]{Secrest_2021}, the number of local sources expected to be flagged for removal should be about $\mathcal{O}(10^4)$, which is greater than the 297 sources flagged for removal when we cross-match with 2MRS \citep[see also the cross-matching in][]{wagenveld2025}. However, considering the incompleteness of 2MRS and that the CatWISE quasars at $z\!<\!0.1$ are effectively at $z\!\sim\!0.1$ rather than $z\!\sim\!0.02$, it is reasonable that, in practice, cross-matching selects \orange{fewer} sources for removal than that estimated. Therefore, the population of low redshift sources remaining in CatWISE is expected to be insignificant and likely cannot \orange{explain} the $>\!5\sigma$ tension with \textit{Planck}.}

\red{In light of the above, it appears unlikely that a} population of unidentified local sources in NVSS, RACS-low and CatWISE is the cause of the excessive dipole amplitudes \red{and the tensions that we observe with \textit{Planck}. For further investigation of redshift cross-matching, we turn to current and near-future source catalogues}. These include the 2MASS Photometric Redshift catalogue \citep[2MPZ;][]{Bilicki_2014}, the optical spectroscopic measurements of the Dark Energy Spectrosopic Instrument \citep[DESI;][]{2019BAAS...51g..57L}, the optical and near-infrared observations of the \textit{Euclid} satellite \citep{2024arXiv240513491E}, the photometric measurements of the Legacy Survey of Space and Time \citep[LSST;][]{Ivezić_2019}, the Square Kilometre Array Observatory \citep[SKAO;][]{Bacon2020} radio surveys, and the Spectro-Photometer for the History of the Universe Epoch of Reionization and ices Explorer \citep[SPHEREx;][]{2014arXiv1412.4872D}.

\red{Ultimately, if a significant bulk of the local structure present in the \orange{radio and infrared} surveys in this work has already been \orange{omitted}, enduring tension with \textit{Planck} under the kinematic interpretation} would indicate a major anisotropy in the large-scale structure of the Universe. Such a finding would be incompatible with the CP that underpins the FLRW metric of spacetime. It would question our modern cosmological view and break the standard $\Lambda$CDM concordance model. As it now stands, the source of the \red{tension between \textit{Planck} and} NVSS, RACS-low and CatWISE is unknown \red{and constitutes a significant outstanding anomaly in cosmology.}

\subsection{Tension Between Dipoles in \orange{the Radio and Infrared} Surveys}\label{sec:internal-tensions}

\begin{figure}
    \includegraphics[width=\columnwidth]{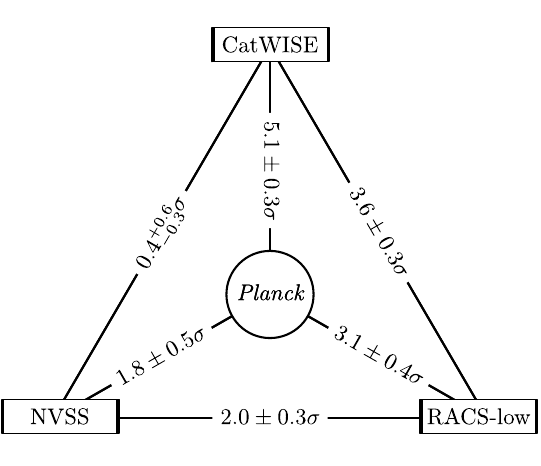}
    \caption{Illustration of the tension between the surveys analysed in this work, under the kinematic interpretation of the dipole. The outside triad \newred{are} the \orange{tensions between the} dipoles inferred in NVSS~--~B, RACS-low~--~B and CatWISE. The interior tensions are with the CMB as observed by \textit{Planck}.}
    \label{fig:triangle}
\end{figure}

Separate \orange{from} the consistency check between \orange{the source count dipole and} the dipole in the CMB, the populations of radio sources and infrared-selected quasars provide an additional test of the distribution of matter, \newred{under} the CP. Fundamentally, \newred{the kinematic dipole inferred from} \red{the source count dipole in} these \red{surveys} should be concordant.

\red{We observe significant} concordance between NVSS and CatWISE, at only \red{$0.4\sigma$} tension under the kinematic interpretation \red{(see also their highly concordant posteriors in Figure \ref{fig:corner-plots}). This corroborates the agreement between NVSS and CatWISE \orange{found} in \citet{Secrest_2022}. These surveys have completely independent systematics, where NVSS is a radio-continuum ground-based survey and CatWISE is derived from near-infrared satellite observations. This indicates that the anomalous dipole present in NVSS and CatWISE is a common signal, arising from some unknown physical mechanism.}

\red{Conversely,} RACS-low is \textit{discordant} with both CatWISE (at about \red{$3.6\sigma$} tension) and NVSS (at about $2\sigma$ tension). \newred{Whilst radio sources selected at different frequencies trace different populations, the populations at $\sim\!900$~MHz and 1.4~GHz are comparable~\citep[see, e.g.][]{2003MNRAS.342.1117M}. It is therefore} curious that \orange{RACS-low is in strong tension with CatWISE whilst NVSS is concordant with CatWISE. Considering} the triad of \orange{radio and infrared survey tensions} (see Figure~\ref{fig:triangle}), RACS-low being in moderate \orange{or strong} tension with all of the datasets used in this work suggests a possible systematic \red{difference} in the catalogue itself, or the underlying observational data. The suite of RACS observations at differing central frequencies, resolution convolutions and flux scalings provides an opportunity to probe this further \citep[see, e.g.][]{Hale_2021,2023PASA...40...34D,2024PASA...41....3D,2025PASA...42...38D}. To \orange{investigate} the consistency of the dipole in \orange{radio} surveys, a comprehensive analysis of the observational and cataloguing rationales is required.

Recall \orange{from Section~\ref{subsec:independant-datasets}} that for valid analysis and Bayesian tensions we require datasets to be independent. The analysis we present here, however, is not wholly independent -- there are common sources in the samples. \orange{Before} analysis, it is reasonable to control for this by cross-matching and removing duplicated sources \citep[see, e.g. the cross-matching of NVSS and CatWISE in][]{Secrest_2022,wagenveld2025}. In practice, however, the different emissions of \orange{AGN in the} populations, and the systematics of each survey, such as differing angular resolution and astrometric errors, cause limitations akin to \orange{those} discussed in Section \ref{sec:discussion-redshift}. \green{Whilst} complete dataset independence is not possible with cross-matching, \red{partial independence is.} \red{Since only $\sim\!2$ percent of WISE-selected AGNs are radio-loud \citep[see, e.g.][]{Stern_2012}, independence between the radio surveys and CatWISE is not a significant limitation. For radio-to-radio, despite the aforementioned resolution and astrometry limitations, most sources in the other catalogue should readily match.} \green{Moreover, dataset independence could be enforced by extending the likelihood function, with survey overlap modelled and marginalised over.}

\green{\orange{Alternatively, one could construct} independent survey footprints through the choice of mask \citep[see, e.g. that in][]{wagenveld2025}.} \red{Here, using the B variants, we mask NVSS below $\delta=0^\circ$ and RACS-low above, resulting in a sample of 205\,586 and 307\,838 sources, respectively. Using the approach from Section~\ref{sec:nested-sampling-approach}, we find that NVSS and RACS-low are in $2.0\substack{+0.5\\-0.4}\sigma$ tension. Masking NVSS below $b=0^\circ$ and RACS-low above, resulting in a sample of 197\,438 and 267\,616 sources, respectively, we find $1.9\substack{+0.4\\-0.3}\sigma$ tension. In both analysis approaches, the surveys are completely independent. The recovered NVSS--RACS tensions remain moderate and are consistent with Figure \ref{fig:triangle}. Therefore, we reaffirm that the analysis here suggests that NVSS and CatWISE share a common signal, and that the RACS-low catalogue may contain an unknown, systematic difference.}

\subsection{Source \red{Count} Dependence of the Tension Between \textit{Planck} and \red{the} Radio Surveys}

\begin{figure}
    \includegraphics[width=\columnwidth]{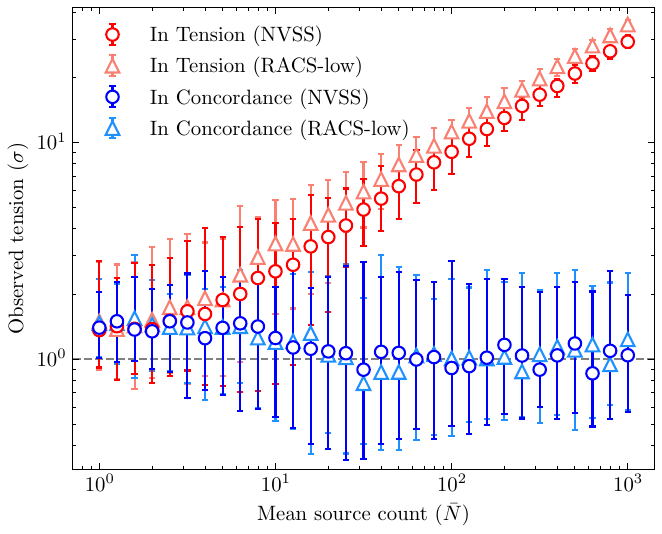} \\ [1ex]
    \includegraphics[width=\columnwidth]{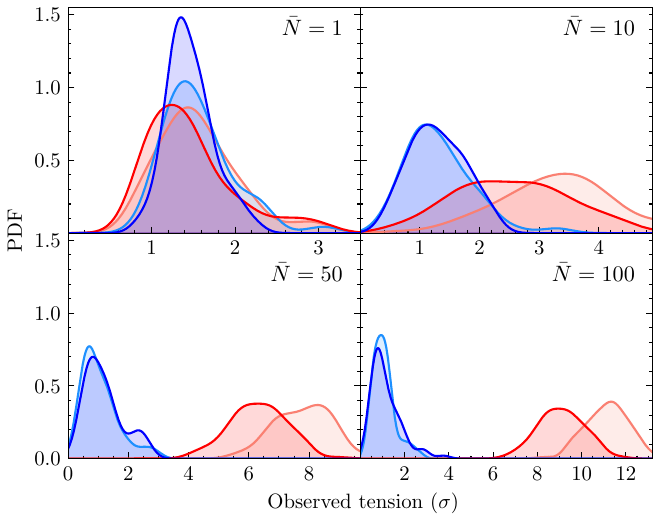}
    \caption{Comparison of the observed tension of the synthetic NVSS and RACS-low samples in tension and in concordance with \textit{Planck}. \textit{Top:} the distribution of the observed tension with \textit{Planck} at each mean source count. The circle or triangle is the median of each distribution, with the $2\sigma$ credible interval given by the error bars. The horizontal dashed grey line marks $1\sigma$ tension. \textit{Bottom:} the observed tension distributions at four mean source count slices from the top panel. The distributions are estimated using a Gaussian kernel, normalised as a probability density function (PDF).}
    \label{fig:tension-vs-multiplier}
\end{figure}

The tension between \textit{Planck} and the radio \orange{samples in this work} is not at the \orange{severe $\sim\!5\sigma$ level that CatWISE is.} \red{Whilst recently a dipole three times the kinematic expectation at $4.8\sigma$ was \orange{jointly} estimated using radio catalogues \citep{Wagenveld_2023}, the constituent datasets are in moderate tension \orange{(recall from} Section \ref{sec:internal-tensions}\orange{)}. \orange{This result is not equivalent to the challenge against the standard model posed by CatWISE, since the joint} analysis of datasets that are in tension \orange{is} not necessarily valid \citep[see, e.g.][]{PhysRevD.100.043504}.} \orange{Therefore, it is prudent to further probe these individual radio survey tensions.}

The posteriors of NVSS and RACS-low, \orange{which are in tension with \textit{Planck}}, are \orange{quite} unconstrained. \orange{If these posteriors were either tighter due to additional information from higher source counts, or centred further away from that of \textit{Planck}, we would observe the tension to increase. These two factors will dictate whether the tension between \textit{Planck} and future radio surveys is trivial or significant.} Here, we investigate how the \orange{total number of sources} affects the \orange{tension with \textit{Planck}, assuming that the posterior centres remain fixed at that of either NVSS or RACS-low.}

We \orange{generate} synthetic samples (mock skies) with \orange{different} total number\orange{s} of sources \orange{by controlling the} mean source count, $\bar{N}$. For 31 \red{logarithmically spaced} values of $\bar{N}$ from 1--$10^3$, we construct synthetic samples of the B variants of NVSS and RACS-low using the Table \ref{table:results-individual} \orange{best-}fits and the Section \ref{sec:expected-amp} expected amplitudes:
\begin{enumerate}
    \renewcommand{\labelenumi}{\roman{enumi}.}
    \item Given some median inferred dipole values $\{v,l,b\}$ and an expected amplitude $\tilde{\mathcal{D}}$, we compute the dipole $\mathbf{d}\!=\!\mathcal{D}\hat{\mathbf{d}}$;
    \item Given the choice of $\bar{N}$, for each healpixel $\hat{\mathbf{n}}_i$, we compute the expectation value $\lambda_i$ and draw from a Poisson distribution using the \textsc{Numpy} function \texttt{random.poisson} with $\lambda\!=\!\lambda_i$ \citep{Harris_2020};
    \item We then measure the tension between the synthetic sample and our actual \textit{Planck} dataset from Section \ref{sec:planck-intro}; and
    \item Finally, we repeat these steps using 100 different seeds.
\end{enumerate}
This series of synthetic NVSS and RACS-low samples \orange{are designed to be in tension with \textit{Planck}, since their}  dipoles differ from the \orange{kinematic} expectation. We repeat the above routine, except with both surveys using the \orange{best-fit median $\{v,l,b\}$ of} \textit{Planck} \orange{from} Table~\ref{table:results-individual}. This second series of synthetic NVSS and RACS-low samples \orange{is designed} to be in concordance with \textit{Planck}. \red{We find that \orange{applying the respective mask of the survey} has no significant effect on the analysis, being \orange{effectively} equivalent to reducing $\bar{N}$. Therefore, we focus on the unmasked synthetic samples.}

\orange{S}ince the synthetic samples were each regenerated 100 times\orange{, the observed tensions form a distribution at each mean source count} (see the top panel of Figure \ref{fig:tension-vs-multiplier}). \red{We find that the observed tension of the in tension series of RACS-low is consistently about 1.2 times that of NVSS. Additionally,} the in tension and in concordance series diverge as the mean source count increases. Above $\bar{N}\!\approx\!3$, the observed tension of the in tension series of NVSS and RACS-low both evolve as approximately $\sigma(\bar{N})\!\propto\!\bar{N}^{1/2}$, \red{in line with the signal-to-noise ratio (SNR) of a Poisson system, which evolves as $\sqrt{N}$. 
Whilst the in concordance series oscillates around $1\sigma$ at high source count, at low source count both series converge to about $1.6\sigma$ \orange{tension}. Here, where the SNR is low and shot noise dominates, the posteriors \orange{of both series }are unconstrained, \orange{causing them to} present as effectively the same.}

\orange{Turning to} the probability density of the distributions, the degeneracy between the in tension and in concordance series becomes clearer (see the bottom panel of Figure \ref{fig:tension-vs-multiplier}). Degeneracy between the in concordance and in tension series indicates an inability to make credible inference due to a low level of information. Whilst $\bar{N}\!=\!1$ is an extreme hypothetical, corresponding to only about 50\,000 sources, there remains substantial degeneracy at $\bar{N}\!=\!10$, which is analogous to the radio samples we use in this work. This indicates that, at our current level of information, our observed \textit{Planck}--NVSS and \textit{Planck}--RACS-low tensions may be due to \orange{intrinsic} disagreement with the CMB, or \orange{due to} shot noise and other systematics. In the regimes of the bottom-most sub-panels, where $\bar{N}\!=\!50$ and 100, we observe tension with \textit{Planck} that is distinct from concordance, which indicates a high ability to make credible inference.

In the future, \orange{higher} source counts will \orange{allow practitioners} to probe the dipole \orange{in radio surveys} in a regime closer to $\bar{N}\!=\!50$ or 100\newred{, with the caveat that one must effectively remove low-$z$ star-forming galaxies, which dominate at low radio fluxes~\citep[see, e.g.][]{2016A&ARv..24...13P}}. Assuming a dipole with the median $\{v,l,b\}$ \red{of the B variant} of NVSS, we predict that the tension with \textit{Planck} will be measured at a significance of $4\sigma$ with about \red{1\,000\,000} sources, or $5\sigma$ with about \red{1\,600\,000} sources. If the dipole is instead that \red{of the B variant} of RACS-low, we predict a $4\sigma$ or $5\sigma$ measurement with about \red{800\,000} or \red{1\,100\,000} sources, respectively. This is about $\mathcal{O}(10^6)$ radio sources, which is well within the scope of near-future surveys.


\green{Whilst} \red{similar calculations performed by \citet{2009ApJ...692..887C} indicate that a catalogue of about \green{two million} is required to detect the kinematic dipole \orange{to} $3\sigma${, the} calculations that we perform here \green{differ \orange{since} they} assume an excessive dipole amplitude. \green{This is in line with the dipoles revealed in the radio samples used \orange{here} and throughout the literature. \orange{With} a \orange{dipole amplitude} exceeding the kinematic expectation by a factor of 2--3, it is intuitive that \orange{fewer} sources can allow for measurements \orange{of higher significance} than that predicted by \citet{2009ApJ...692..887C}. Given our prediction of $\mathcal{O}(10^6)$ \orange{radio} sources, we are on the cusp of significant tension measurements in individual \orange{analyses of} radio surveys that would firmly add to the challenge against the standard model already posed by infrared-selected quasars.}}


\subsection{Future Outlook}
\green{\orange{Whilst} the tension investigation in this work analyses datasets in pairs,} the approach allows for datasets to be included ad infinitum, remaining valid so long as the dipole in each is induced by \orange{the same} mechanism, as assumed by the model. The kinematic interpretation of the dipole is such an assumption. \green{Therefore, future work may include a tension investigation of greater than two datasets at once.}

\orange{Additionally, tension investigations may be extended to include higher-order multipoles in the underlying model \citep[see, e.g.][]{10.1093/mnras/stae2776}. However, care must be taken when choosing which higher-order signals to treat as common (from astrophysical sources) or nuisance (from systematics), which is an unsolved problem.}



Ultimately, we will see substantial developments in inference with the additional observations of millions of cosmologically distant objects in upcoming surveys. The LSST will observe about $10^7$ quasars \citep{Ivezić_2019}, the Euclid Wide Survey \citep[EWS;][]{2022A&A...662A.112E} will observe about $4\!\times\!10^7$ active galactic nuclei, and the SKA will observe about $5\!\times\!10^6$ galaxies in Phase 1 and $9\!\times\!10^8$ galaxies in Phase 2 \citep{10.1093/mnras/stv040,Bacon2020}. With the approach in this work, these massive cosmological datasets are a profound opportunity to significantly measure the dipole tension and decisively reconcile the cosmic dipole anomaly.

\section*{Acknowledgements}
We thank Will Handley for insightful discussions on Bayesian ten-
sions and the anonymous referee for their helpful report. We also
thank Oliver Oayda for useful discussions regarding the NVSS and
RACS-low samples. Additionally, we thank Emil Lenc and Stefan
Duchesne for their helpful comments on the RACS samples. MLS
was supported by the Australian Government Research Training Pro-
gram (RTP) Scholarship.

This work made use of the \textit{Planck} Public Data Release 3 \citep{2020ipac.data.I558P}, the \textsc{BeyondPlanck} Data Release II \citep{2023A&A...675A...1B}, the National Radio Astronomy Observatory Very Large Array Sky Survey \citep{Condon_1998}, the Rapid Australian Square Kilometre Array Pathfinder Continuum Survey \citep{McConnell_2020}, the CatWISE2020 data release \citep{Eisenhardt_2020}, the Two Micron All Sky Survey Redshift Survey \citep{Huchra_2012}, and the NASA/IPAC Extragalactic Database.


This scientific work uses data obtained from Inyarrimanha Ilgari Bundara, the CSIRO Murchison Radio-astronomy Observatory. We acknowledge the Wajarri Yamaji People as the Traditional Owners and native title holders of the Observatory site. CSIRO’s ASKAP radio telescope is part of the Australia Telescope National Facility (\url{https://ror.org/05qajvd42}). Operation of ASKAP is funded by the Australian Government with support from the National Collaborative Research Infrastructure Strategy. ASKAP uses the resources of the Pawsey Supercomputing Research Centre. Establishment of ASKAP, Inyarrimanha Ilgari Bundara, the CSIRO Murchison Radio-astronomy Observatory and the Pawsey Supercomputing Research Centre are initiatives of the Australian Government, with support from the Government of Western Australia and the Science and Industry Endowment Fund.

This work used the \textsc{Python} packages \textsc{Anesthetic} \citep{anesthetic}, \textsc{Astropy} \citep{2022ApJ...935..167A}, \textsc{Healpy} \citep{2005ApJ...622..759G,Zonca2019}, \textsc{Matplotlib} \citep{4160265}, \textsc{Numpy} \citep{Harris_2020}, \textsc{Pandas} \citep{mckinney-proc-scipy-2010,reback2020pandas}, \textsc{Scipy} \citep{2020NatMe..17..261V}, and \textsc{Ultranest} \citep{2021JOSS....6.3001B}.


\section*{Data Availability}
The ASKAP data used in this work is publicly available from the CSIRO
ASKAP Science Data Archive (CASDA; \url{https://research.csiro.au/casda}) under project code AS110. Any additional data products used in this work are publicly available online. The samples used in this study will be made available upon reasonable request to the authors.
 



\bibliographystyle{mnras}







\bsp	
\label{lastpage}
\end{document}


\label{firstpage}
\pagerange{\pageref{firstpage}--\pageref{lastpage}}
\maketitle


\begin{keywords}
errata, addenda -- large-scale structure of Universe -- cosmology: observations -- cosmology: theory -- radio continuum: galaxies -- infrared: galaxies -- cosmic background radiation
\end{keywords}



\begin{figure*}
    \centering
        \includegraphics[width=\columnwidth]{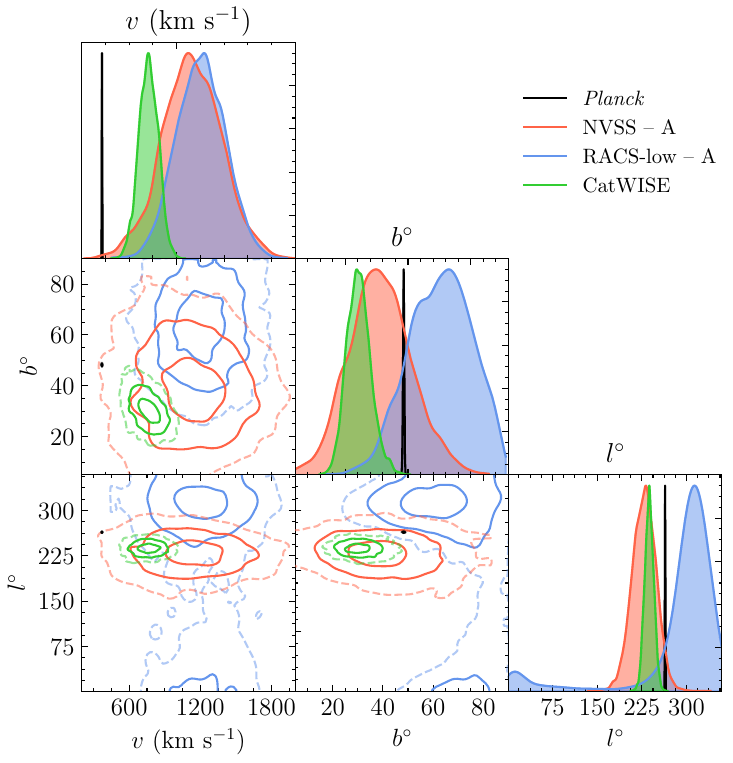}
    \hfill
        \includegraphics[width=\columnwidth]{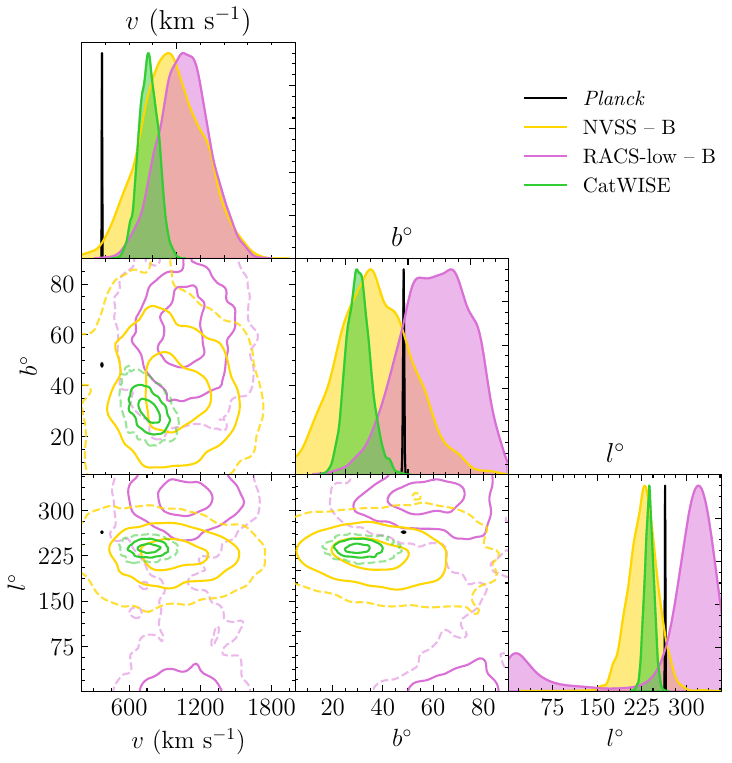}
    \caption{Posteriors from the individual analyses of \textit{Planck}, NVSS, RACS-low and CatWISE. The nuisance parameters are omitted for clarity. The contours of the 2D marginal posteriors are in intervals of $1\sigma$, encompassing 39.4, 86.4 and 98.9 (dashed) percent of the probability distribution. These contours have been smoothed by three percent using a Gaussian kernel. \textit{Left:} the posteriors of \textit{Planck}, CatWISE and the A variants of NVSS and RACS-low. \textit{Right:} the posteriors of \textit{Planck}, CatWISE and the B variants of NVSS and RACS-low.}
    \label{fig:corner-plots}
\end{figure*}

\begin{figure}
    \includegraphics[width=\columnwidth]{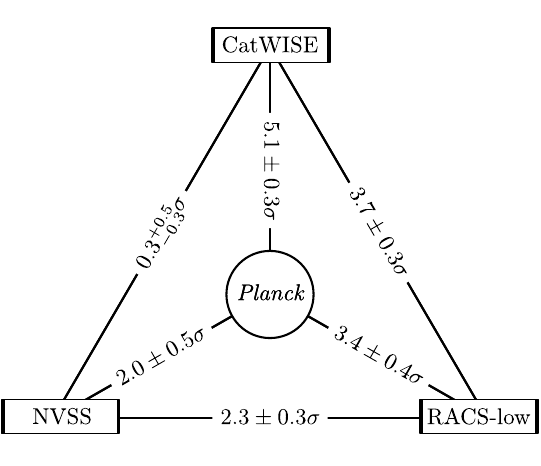}
    \caption{Illustration of the tension between the surveys analysed in this work, under the kinematic interpretation of the dipole. The outside triad \newred{are} the \orange{tensions between the} dipoles inferred in NVSS~--~B, RACS-low~--~B and CatWISE, and the interior tensions are with \textit{Planck}. The errors quoted here are $2\sigma$ errors.}
    \label{fig:triangle}
\end{figure}

\begin{figure}
    \includegraphics[width=\columnwidth]{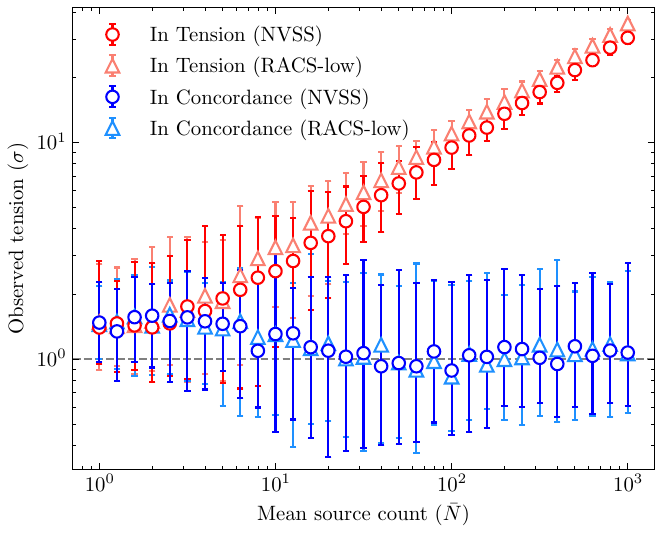} \\ [1ex]
    \includegraphics[width=\columnwidth]{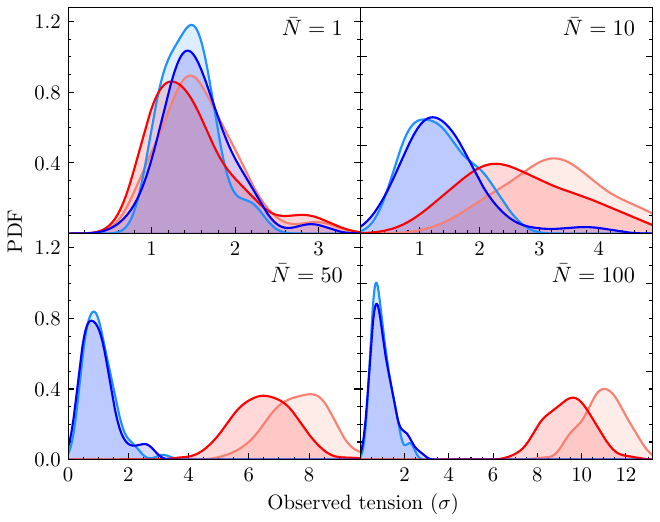}
    \caption{Comparison of the observed tension of the synthetic NVSS and RACS-low samples in tension and in concordance with \textit{Planck}. \textit{Top:} the distribution of the observed tension with \textit{Planck} at each mean source count. The circle or triangle is the median of each distribution, with the $2\sigma$ credible interval given by the error bars. The horizontal dashed grey line marks $1\sigma$ tension. \textit{Bottom:} the observed tension distributions at four mean source count slices from the top panel. The distributions are estimated using a Gaussian kernel, normalised as a probability density function (PDF).}
    \label{fig:tension-vs-multiplier}
\end{figure}

In the original paper \citep[][]{10.1093/mnras/staf1621}, incorrectly a Gaussian rather than a Poisson function was used in the likelihood. This resulted in small changes to tables 1 and 2, figures 3 to 5, and sections 6.3 and 6.4, which we present here for completeness (see Tables~\ref{table:results-individual} and~\ref{table:results-joint}, and Figures~\ref{fig:corner-plots}, \ref{fig:triangle} and~\ref{fig:tension-vs-multiplier}). The conclusions of the original paper remain unchanged.

Correction to the last paragraph of section~6.3 in \citet{10.1093/mnras/staf1621}: \green{\orange{Alternatively, one could construct} independent survey footprints through the choice of mask \citep[see, e.g. that in][]{wagenveld2025}.} \red{Here, using the B variants, we mask NVSS below $\delta=0^\circ$ and RACS-low above, resulting in a sample of 205\,586 and 307\,838 sources, respectively. Using the approach from section~4 in \citet{10.1093/mnras/staf1621}, we find that NVSS and RACS-low are in $2.3\substack{+0.5\\-0.4}\sigma$ tension. Masking NVSS below $b=0^\circ$ and RACS-low above, resulting in a sample of 197\,438 and 267\,616 sources, respectively, we find $2.2\substack{+0.4\\-0.3}\sigma$ tension. In both analysis approaches, the surveys are completely independent. The recovered NVSS--RACS tensions remain moderate and are consistent with Figure \ref{fig:triangle}. Therefore, we reaffirm that the analysis here suggests that NVSS and CatWISE share a common signal, and that the RACS-low catalogue may contain an unknown, systematic difference.}

{\parfillskip=0pt
Correction to the second-last paragraph of section~6.4 in \citet{10.1093/mnras/staf1621}: In the future, \orange{higher} source counts will \orange{allow practitioners} to probe the dipole \orange{in radio surveys} in a regime closer to $\bar{N}\!=\!50$ or 100\newred{, with the caveat that one must effectively remove low-$z$ star-forming galaxies, which dominate at low radio fluxes~(see,}\par}

\vfill\null\newpage


\noindent e.g. \citealt{2016A&ARv..24...13P}). Assuming a dipole with the median $\{v,l,b\}$ \red{of the B variant} of NVSS, we predict that the tension with \textit{Planck} will be measured at a significance of $4\sigma$ with about \red{1\,000\,000} sources, or $5\sigma$ with about \red{1\,500\,000} sources. If the dipole is instead that \red{of the B variant} of RACS-low, we predict a $4\sigma$ or $5\sigma$ measurement with about \red{800\,000} or \red{1\,200\,000} sources, respectively. This is about $\mathcal{O}(10^6)$ radio sources, which is well within the scope of near-future surveys.

\begin{table}
\centering
\begin{tabular}{@{\hskip 1pt}l@{\hskip 4.5pt}c@{\hskip 4.5pt}c@{\hskip 4.5pt}c@{\hskip 4.5pt}c@{\hskip 4.5pt}c@{\hskip 4.5pt}c@{\hskip 4.5pt}c@{\hskip 1pt}}
\hline \\ [-3ex]
 &&\multicolumn{3}{c}{Inferred Dipole}&\multicolumn{2}{c}{Nuisance Parameters}\\
 \hline \\ [-3ex]
 Dataset&&$v$ [km $\text{s}^{-1}$]&$l~[^\circ]$&$b~[^\circ]$&$\bar{N}~[\text{px}\substack{-1\\{}}]$&$\Upsilon_\text{ecl}$\\
 \hline \\ [-2.7ex]
 \textit{Planck}&--&$370\!\pm\!3$&${264.1}_{-0.7}^{+0.8}$&${48.2}\!\pm\!{0.5}$&${2725498}_{-17}^{+16}$&--\\[1ex]
 NVSS&A&${1126}_{-514}^{+500}$&${229}_{-41}^{+37}$&${39}_{-23}^{+26}$&${10.16}_{-0.03}^{+0.04}$&--\\[1ex]
 "&B&${948}_{-503}^{+526}$&${228}_{-51}^{+46}$&${36}_{-28}^{+31}$&${10.07}\!\pm\!{0.04}$&--\\[1ex]
 RACS-low&A&${1207}_{-400}^{+399}$&${310}_{-300}^{+43}$&${64}_{-25}^{+21}$&${14.17}\!\pm\!{0.05}$&--\\[1ex]
 "&B&${1062}_{-407}^{+403}$&${312}_{-307}^{+42}$&${62}_{-28}^{+22}$&${14.06}_{-0.04}^{+0.05}$&--\\[1ex]
 CatWISE&--&${763}_{-162}^{+156}$&${238}\!\pm\!{15}$&${30}_{-9}^{+10}$&${68.6}\!\pm\!{0.2}$&${0.99}_{-0.08}^{+0.07}$\\
 \hline
\end{tabular}
\caption{The inferred dipole and nuisance parameters of the individual analyses of \textit{Planck}, NVSS, RACS-low and CatWISE. The value in each cell is the median of its respective marginalised posterior, with the upper and lower bounds of its $2\sigma$ credible interval. The $\bar{N}$ parameter is in units of counts per healpixel, where the counts are \textmu K for \textit{Planck} and sources for NVSS, RACS-low and CatWISE.}\label{table:results-individual}
\end{table}

\begin{table*}
\centering
\begin{tabular}{@{\hskip 1.5pt}l@{\hskip 5pt}c@{\hskip 5pt}c@{\hskip 5pt}c@{\hskip 5pt}c@{\hskip 5pt}c@{\hskip 5pt}c@{\hskip 5pt}c@{\hskip 5pt}c@{\hskip 5pt}c@{\hskip 5pt}c@{\hskip 5pt}c@{\hskip 5pt}c@{\hskip 1.5pt}}
\hline \\ [-3ex]
 &&\multicolumn{3}{c}{Inferred Dipole}&\multicolumn{3}{c}{Nuisance Parameters}&\multicolumn{5}{c}{Bayesian Statistics}\\
 \hline \\ [-3ex]
 Dataset&&$v$ [km $\text{s}^{-1}$]&$l~[^\circ]$&$b~[^\circ]$&$\bar{N}_1~[\text{px}\substack{-1\\{}}]$&$\bar{N}_2~[\text{px}\substack{-1\\{}}]$&$\Upsilon_\text{ecl}$&$\log{R}$&$\log{I}$&$\log{S}$&$d$&$\sigma$\\
 \hline \\ [-2.7ex]
 \textit{Planck}--NVSS&A&${370}\!\pm\!{3}$&${264.1}\!\pm\!{0.7}$&${48.2}\!\pm\!{0.5}$&${2725497}\!\pm\!{17}$&${10.16}\!\pm\!{0.03}$&--&${0.2}\!\pm\!{0.8}$&${4.6}\!\pm\!{0.8}$&${-4.4}\!\pm\!{1.1}$&${3.1}_{-0.5}^{+0.6}$&${2.6}\!\pm\!{0.4}$\\[1ex]
 "&B&${370}\!\pm\!{3}$&${264.1}\!\pm\!{0.8}$&${48.3}\!\pm\!{0.5}$&${2725498}\!\pm\!{17}$&${10.08}\!\pm\!{0.03}$&--&${1.2}\!\pm\!{0.8}$&${3.5}\!\pm\!{0.8}$&${-2.3}\!\pm\!{1.1}$&${2.6}_{-0.5}^{+0.6}$&${2.0}\!\pm\!{0.5}$\\[1ex]
 \textit{Planck}--RACS-low&A&${370}\!\pm\!{3}$&${264.1}\!\pm\!{0.8}$&${48.2}\!\pm\!{0.5}$&${2725497}_{-17}^{+18}$&${14.16}\!\pm\!{0.04}$&--&${-4.4}\!\pm\!{0.8}$&${4.8}\!\pm\!{0.8}$&${-9.1}\!\pm\!{1.1}$&${2.5}_{-0.5}^{+0.6}$&${4.0}\!\pm\!{0.4}$\\[1ex]
 "&B&${370}\!\pm\!{3}$&${264.1}\!\pm\!{0.8}$&${48.2}\!\pm\!{0.5}$&${2725497}_{-16}^{+18}$&${14.05}\!\pm\!{0.04}$&--&${-1.4}\!\pm\!{0.8}$&${5.2}\!\pm\!{0.8}$&${-6.6}\!\pm\!{1.1}$&${2.4}_{-0.5}^{+0.6}$&${3.4}\!\pm\!{0.4}$\\[1ex]
 \textit{Planck}--CatWISE&--&${370}\!\pm\!{3}$&${264.0}_{-0.7}^{+0.8}$&${48.2}\!\pm\!{0.5}$&${2725497}_{-17}^{+18}$&${68.6}\!\pm\!{0.2}$&${0.99}_{-0.07}^{+0.08}$&${-6.4}\!\pm\!{0.9}$&${8.4}\!\pm\!{0.8}$&${-14.7}\!\pm\!{1.2}$&${3.0}_{-0.6}^{+0.7}$&${5.1}\!\pm\!{0.3}$\\[1ex]
 NVSS--RACS-low&A&${1055}_{-290}^{+304}$&${267}_{-59}^{+40}$&${61}_{-21}^{+19}$&${10.16}\!\pm\!{0.04}$&${14.17}_{-0.04}^{+0.05}$&--&${0.3}\!\pm\!{0.5}$&${4.0}\!\pm\!{0.5}$&${-3.7}\!\pm\!{0.7}$&${2.7}_{-0.5}^{+0.6}$&${2.5}\!\pm\!{0.3}$\\[1ex]
 "&B&${905}_{-297}^{+299}$&${269}_{-89}^{+59}$&${62}_{-24}^{+20}$&${10.07}_{-0.04}^{+0.03}$&${14.06}\!\pm\!{0.04}$&--&${0.6}\!\pm\!{0.5}$&${3.8}\!\pm\!{0.5}$&${-3.3}\!\pm\!{0.7}$&${2.8}_{-0.5}^{+0.6}$&${2.3}\!\pm\!{0.3}$\\[1ex]
 NVSS--CatWISE&A&${792}_{-149}^{+152}$&${237}_{-15}^{+13}$&${30}_{-8}^{+10}$&${10.16}\!\pm\!{0.03}$&${68.6}\!\pm\!{0.2}$&${0.99}_{-0.08}^{+0.07}$&${4.7}\!\pm\!{0.6}$&${4.8}\!\pm\!{0.6}$&${-0.1}\!\pm\!{0.9}$&${3.6}_{-0.6}^{+0.7}$&${0.9}\!\pm\!{0.5}$\\[1ex]
 "&B&${776}_{-145}^{+153}$&${237}\!\pm\!{14}$&${31}_{-8}^{+10}$&${10.08}\!\pm\!{0.03}$&${68.6}\!\pm\!{0.2}$&${0.99}\!\pm\!{0.07}$&${5.0}\!\pm\!{0.6}$&${4.1}\!\pm\!{0.6}$&${0.9}\!\pm\!{0.9}$&${3.0}_{-0.6}^{+0.7}$&${0.3}_{-0.3}^{+0.5}$\\[1ex]
 RACS-low--CatWISE&A&${727}_{-141}^{+144}$&${245}_{-17}^{+16}$&${35}_{-9}^{+11}$&${14.16}\!\pm\!{0.04}$&${68.6}\!\pm\!{0.2}$&${0.99}\!\pm\!{0.08}$&${-4.6}_{-0.7}^{+0.6}$&${5.2}\!\pm\!{0.6}$&${-9.9}\!\pm\!{0.9}$&${2.9}_{-0.6}^{+0.7}$&${4.1}\!\pm\!{0.3}$\\[1ex]
 "&B&${711}_{-136}^{+144}$&${244}\!\pm\!{16}$&${36}_{-9}^{+11}$&${14.05}\!\pm\!{0.04}$&${68.6}\!\pm\!{0.2}$&${0.99}\!\pm\!{0.07}$&${-2.8}\!\pm\!{0.6}$&${5.5}\!\pm\!{0.6}$&${-8.4}\!\pm\!{0.9}$&${3.2}_{-0.6}^{+0.7}$&${3.7}\!\pm\!{0.3}$\\
 \hline
\end{tabular}
\caption{The inferred dipole, nuisance parameters and Bayesian statistics of the joint analyses of \textit{Planck}, NVSS, RACS-low and CatWISE. The value in each cell is the median of its respective marginalised posterior, with the upper and lower bounds of its $2\sigma$ credible interval. The $\bar{N}_1$ and $\bar{N}_2$ parameters are in units of counts per healpixel, where the counts are \textmu K for \textit{Planck} and sources for NVSS, RACS-low and CatWISE.}\label{table:results-joint}
\end{table*}

\null\newpage
\null\newpage
\null\newpage
\null\newpage



\bibliographystyle{mnras}







\bsp	
\label{lastpage}